\documentclass[12pt]{article}
\usepackage{amssymb,amsmath}
\usepackage{psfig}
\begin{document}

\title{Quantum Resonances and Regularity Islands in Quantum Maps}
\author{V.V. Sokolov \thanks{E-mail: vsokolov@inp.nsk.su},
O.V. Zhirov \thanks{E-mail: zhirov@inp.nsk.su} \\
\it Budker Institute of Nuclear Physics, 630090, Novosibirsk-90,
Russia \\
D. Alonso \thanks{E-mail: dalonso@ull.es},\\
 \it Departamento de Fisica Fundamental
y Experimental, \\
\it La Laguna 38203, Tenerife, Spain \\
 and G. Casati
\thanks{E-mail: casati@fis.unico.it},\\
\it International Center for the Study of Dynamical Systems, 22100
Como,\\
\it and Istituto Nazionale di Fisica della Materia and INFN, \\
\it Unita'di Milano, Italy}

\date{\today}

\maketitle

\begin{abstract}
We study analytically as well as numerically the dynamics of a
quantum map near a quantum resonance of an order $q$. The map is
embedded into a continuous unitary transformation generated by a
time-independent quasi-Hamiltonian. Such a Hamiltonian generates at
the very point of the resonance a local gauge transformation
described the unitary unimodular group $SU(q)$. The resonant energy
growth of is attributed to the zero Liouville eigenmodes of the
generator in the adjoint representation of the group while the
non-zero modes yield saturating with time contribution. In a
vicinity of a given resonance, the quasi-Hamiltonian is then found
in the form of power expansion with respect to the detuning from
the resonance. The problem is related in this way to the motion
along a circle in a $(q^2-1)$-component inhomogeneous ``magnetic"
field of a quantum particle with $q$ intrinsic degrees of freedom
described by the $SU(q)$ group. This motion is in parallel with the
classical phase oscillations near a non-linear resonance. The most
important role is played by the resonances with the orders much
smaller than the typical localization length, $q\ll l$. Such
resonances master for exponentially long though finite times the
motion in some domains around them. Explicit analytical solution is
possible for a few lowest and strongest resonances.\\
[2ex]
\end{abstract}

\setcounter{equation}{0}
\section{Introduction}

Classical canonical two-dimensional maps originated from the
Poincare sections in the phase space has played an exceptional role
in the establishing of our understanding of the origin and
properties of the dynamical chaos \cite{Chir79,LL83}. Formally,
they correspond to non-conservative Hamiltonian systems with one
degree of freedom driven by instantaneous periodic kicks. The phase
plane of such a map is, generally, very complex and consists of
intimately entangled domains filled by regular and chaotic
trajectories. The chaotic domains remain isolated from each other
if the driven force is weak, but they join and global chaos appears
when the strength of the force exceeds some critical value . After
that, regular motion survives only inside isolated islands of the
phase plane, where phase oscillations near the points of mainly low
order non-linear resonances take place, whose areas diminish with
the strength growing.

Existence of the chaotic domains signifies absence of a global
analytical integral of the motion. At the same time, there exist in
the regions of the regular motion approximate local integrals which
can be chosen in many different ways. It should seem, however, that
the most convenient and physically sounding choice is that of the
quasienergy integral which is directly linked to the periodicity of
the driving force. In such an approach \cite{Sok67}, the regular
regime near a non-linear resonance is juxtaposed with the
continuous evolution described by a conservative effective
quasi-Hamiltonian function. Corresponding canonical Hamilton
equations generate continuous trajectories on which all phase
points of the original map lie. The quasi-Hamiltonian is found by
perturbation expansion with respect to a small parameter which can,
in particular, be the closeness to the resonance.

After the quantum extension of the canonical maps had been
suggested in \cite{CCFI79}, the quantum maps were widely used as
informative models of quantum chaos. Amongst them Chirikov's
standard map, i.e. the periodically driven planar rotor,  proved to
be the most economic, fruitful and popular. The unitary Floquet
transformation $U$ which evolves  the QKR wave function
$\psi(\theta)$ over each kick period is given by:
\begin{equation}\label{U}
U = U_r\cdot U_k\equiv\exp\left(-\frac{i}{2}T {\hat m}^2\right)\cdot
\exp\left(-ik\cos\theta\right)
\end{equation}
and consists of successive kick transformation $U_k$ with the
strength $k$ and a free rotation $U_r$ during the time $T$. Here
${\hat m}= -id/d\theta$ and we put $\hbar=1$. The standard map
provides a local description for a large class of dynamical
systems. In particular, there exists a tight and remarkable analogy
\cite{FFGP85} between discovered in \cite{CCFI79} dynamical
suppression of chaos in QKR and Anderson localization in quasi 1D
disordered wires. The diffusive growth of the QKR energy turns
out to be restricted to a certain maximal value because of
dynamical localization in the angular momentum space.

At the same time, there exit some important features of the QKR
dynamics, namely so called quantum resonances, which have no
counterparts in the disordered systems (see \cite{AZ96} and
discussion in \cite{CIS98,AZ98}). At a fixed value of the kick
parameter $k$, special resonant regimes of motion appear
\cite{CCFI79} for everywhere dense set of the rational values
$\varsigma=T/4\pi= p/q$ where the integers $p$ and $q$ are mutually
prime. Under these conditions the rotator regularly accumulates
energy which grows quadratically in the time asymptotics
\cite{IzSh79,Izr90}. Both the restricted diffusion and the quantum
resonances  were experimentally observed in the atom optics
imitation of the QKR reported in \cite{MRBSR95}. Quite recently,
the regime of quantum resonances re-appeared in a new aspect in
connection with the electron scattering with excitation of the
Wannier-Stark resonances \cite{GKK99a,GKK99b}.

 Our
results have already drown much interest of the authors (private
communication).

The interesting and important problem of the impact of the quantum
resonances on the QKR dynamics and the interplay between resonant
and diffusive regimes is still far from a satisfactory
understanding. Investigation of this problem is the main goal of
the present paper. We develop a general approach to the problem of
the motion in a vicinity of a quantum resonance with an arbitrary
{\it order} $q$. Generally, the influence of a quantum resonance
depends on the relation between the order $q$ and the localization
length $l$ in the angular momentum space. We show that the most
important role is played by the resonances with $q\ll l$: for
finite regions around them the quantum motion is explicitly shown
to be regular and dominates the motion for all  values $\varsigma$
inside these regions - the resonance {\it widths}. More precisely,
the motion is well described, during large though finite times, by
a time-independent effective quasi-Hamiltonian with one rotational
degree of freedom and with a discrete spectrum. Such a motion is in
parallel with the classical phase oscillations near a non-linear
resonance. On the contrary, when the resonance order is large
enough, $q\gg l$ the resonant quadratic growth appears only in the
remote time asymptotic and for lesser times the motion reveals
universal features characteristic of the localized quantum chaos.

In sec. 2 we explain the concept of the effective quasi-Hamiltonian
on which our approach is based. The power expansion of the
quasi-Hamiltonian near a quantum resonance is constructed in sec.
3. In this section we also consider analytically and numerically
two strongest {\it boundary} quantum resonances with $q=1, 2$ and
their classical limits. Two more strong resonances are investigated
in the next sec. 4. Contrary to the boundary resonances, they
disappear in the formal limit $\hbar\rightarrow 0$. General
consideration of a resonance of an arbitrary order is presented in
sec. 5. At last, the problem of convergence of our expansion is
discussed in sec. 6.

\section{Quasi-Hamiltonian}
Evolution of the QKR wave function for $n$ kicks is given by $n$
successive repetitions $U(n)\equiv U^n$ of the Floquet
transformation (\ref{U}). Being unitary, the latter can be
expressed in terms of a hermitian operator $\mathcal{H}$ as
$U=\exp(-i\mathcal{H})$. Let us now consider {\it continuous}
unitary transformation $U(t)=\exp(-i\mathcal{H}\,t)$. According to
such a definition, the wave function
$\psi(\theta;t)=U(t)\psi(\theta;0)$ at the integer moments $t=n$
coincides with the quantum state of the map (\ref{U}). On the other
hand, the function $\psi(\theta;t)$ satisfies standard
Schr\"odinger equation with the time-independent Hamiltonian
$\mathcal{H}$. Obviously, the very possibility of the formal
construction described is based on the periodicity of the map.
That is why we refer below to this operator as the {\it
quasi-Hamiltonian}.

Let $|\epsilon\rangle$ be the eigenvector of the Floquet operator
(\ref{U}), which belongs to an eigenvalue $e^{-i\epsilon}$. Then
the quasi-Hamiltonian can be expressed as
\begin{equation}\label{epsH}
\mathcal{H}=\sum_{\epsilon} |\epsilon\rangle\,\epsilon\,\langle\epsilon|
\end{equation}
where the sum runs over the quasienergy spectrum $\{\epsilon\}$ of
the rotator. As usual, each quasienergy is defined up to a term
multiple $2\pi$ which results in corresponding ambiguity of the
quasi-Hamiltonian (\ref{epsH}). However, each time one can fix the
operator $\mathcal{H}$ in the way most convenient for calculation.
The ambiguity does not influence the evolution operator $U(t)$ at
integer moments. As a rule, to get rid of the ambiguity we suggest
continuity of the quasi-Hamiltonian with respect to parameters
under consideration. In the coordinate representation, the
quasi-Hamiltonian is an operator function of the pair of
canonically conjugate observables $\theta$ and ${\hat m}$. Since
the operators ${\hat m}^2$ and $\cos\theta$ do not belong to any
finite algebra, the operator $\mathcal{H}$ cannot generally be
found in a closed form. Rather, it is expressed as an infinite sum
of successive commutators. More than that, one anticipates
extremely non-uniform dependence of the operator $\mathcal{H}$ on
the parameters $\varsigma$ and $k$.

However, the problem simplifies enormously if the condition of a
quantum resonance is fulfilled. As it has been shown by F. Izrailev
and D. Shepelyansky \cite{IzSh79}, at the point of a quantum
resonances with the order $q$ the Floquet operator can, generally,
be presented as a $q\times q$ matrix. This matrix turns out to be
(see below) a local gauge transformation from the unitary $SU(q)$
group generated by a hermitian matrix
$\tilde{\mathcal{H}}^{(res)}(\theta)$ which depends only on the
angle and does not contain the angular momentum. Owing to this
fact, the problem of calculation of the matrix
$\tilde{\mathcal{H}}^{(res)}$ becomes as simple (or complicated) as
that of diagonalization of a $q$-dimensional unitary matrix. The
latter can be carried out analytically if the matrix order does not
exceed 4.

Dependence of the quasi-Hamiltonian on the angular momentum
recovers out of the points of quantum resonances. In some domain
near a given quantum resonance with an order $q$ this dependence
can be found in the form of a power expansion over the detuning
from the resonance. This expansion turns out to appear in the form
of series, in particular, in powers of the angular momentum (see
eq. (\ref{cal Hq}) below) with the resonance matrix
$\tilde{\mathcal{H}}^{(res)}(\theta)$ being the zero-order term in
the series. Actually, such an expansion is quite a formal one. The
question of convergence by no means is trivial. At best, the series
is of only asymptotic nature. Nevertheless, we shall see below that
a few its first terms give surprisingly good description of the
evolution during a very long time.

\section{The Boundary Resonances}
\subsection{Regularity Domains and Quasi-Hamiltonians}

In the simplest case $q=1$, the rotation operator $U_r$ is
equivalent to the identity and the QKR evolution during a time $t$
is described by the unitary phase transformation $e^{-ivt};\,
v(\theta)=k\cos\theta$. This transformation parametrically depends
on the angle $\theta$ and therefore has continuous eigenvalue
spectrum $\epsilon(\theta)=v(\theta)$. By the moment $t$, an
isotropic initial state, which we suggest throughout the paper,
evolves into the wave function
\begin{equation}
\psi(\theta;t)=\frac{1}{\sqrt{2\pi}}\exp(-ikt\cos\theta)
\end{equation}
with $\sim kt$ harmonics. The natural probe of the number of
harmonics is the angular momentum operator ${\hat m}=-id/d\theta$
whose time evolution obeys the linear law
\begin{equation}\label{m1(t)}
{\hat m}(t)\equiv
e^{ivt}\,{\hat m}\,e^{-ivt} = {\hat m} +
e^{ivt}\left[{\hat m}, e^{-ivt}\right]_{-} = {\hat m} -v'\,t; \quad
v'\equiv d v/d\theta.
\end{equation}
This yields the quadratic growth of the kinetic energy $E(t)$ of the
rotor
\begin{equation}\label{varE1}
E(t)\equiv \frac{1}{2}\langle\left[{\hat m}(t)-{\hat m}\right]^2\rangle
= \frac{1}{2}\langle (v')^2\rangle\,t^2 = \frac{k^2}{4}\,t^2
\end{equation}
with the {\it resonant growth rate}$\,\,$ $r=k^2/4$. Here and below
the angular brackets denote averaging over the angle $\theta$.

Let us now consider a vicinity $\kappa=T-4\pi p$ of a main
resonance $q=1$. The time-independent quasi-Hamiltonian ${\cal H}$
in the vicinity is introduced by representing the Floquet operator
in the form
\begin{equation}\label{U1}
U_{p,1}(\kappa) = \exp\left(-\frac{i}{2}\kappa\, {\hat m}^2\right)\exp(-ivt)=
\exp\left(-\frac{i}{\kappa} {\cal H}_1(\kappa)\right)
\end{equation}
with
\begin{equation}\label{cal H1}
{\cal H}_1(\kappa) = \kappa v + \kappa^2 Q(\kappa).
\end{equation}
It follows that the operator $Q(\kappa)$ must satisfy the condition
\begin{equation}\label{Q1}
\exp\left(-\frac{i}{2} \kappa\,{\hat m}^2\right) =
T^*\exp\left\{-i\kappa\int_0^1 d\tau\, Q(\kappa;-\tau)\right\}, \qquad
Q(\kappa;-\tau) = e^{-iv\tau}\,Q(\kappa)\,e^{iv\tau}
\end{equation}
where the symbol $T^*$ stands for the anti-chronological ordering.
Suggesting the operator $Q(\kappa)$ to permit expanding in power
series $Q(\kappa)=Q^{(0)}+\kappa Q^{(1)}+...$ over the detuning
$\kappa$, one comes to successive relations
\begin{equation}\label{eqQ1^0}
\int_0^1 d\tau Q^{(0)}(-\tau) = \frac{1}{2}\,{\hat m}^2,
\end{equation}
\begin{equation}\label{eqQ1^1}
\int\limits^1_0 d\tau Q^{(1)}(-\tau)=
-\frac{i}{2}\int\limits^1_0 d\tau_1\int\limits^{\tau_1}_0 d\tau_2
\left[Q^{(0)}(-\tau_1),Q^{(0)}(-\tau_2)\right]_{-}
\end{equation}
and so on, which allow to find the quasi-Hamiltonian (\ref{cal H1})
up to desired accuracy. Eq. (\ref{eqQ1^0}) implies that the
evolution described by the map (\ref{U}) is smoothed in such a way
that the overall effect of the lowest order correction $\kappa^2
Q^{(0)}$ within one period is identical to the kinetic energy
operator ${\cal K}=J^2/2;\,\, \kappa {\hat m}\equiv J$.
Being a small factor in front of the angle derivative, the detuning
$\kappa$ plays here the role of the dimensionless Planck's constant
while $J$ is the angular momentum operator in the units chosen.
With only the two first corrections (\ref{eqQ1^0},\ref{eqQ1^1})
being retained the quasi-Hamiltonian acquires the form
\begin{equation}\label{apH1} {\cal H}_1 = \frac{1}{2}JF_2(\theta)J +
\frac{1}{2}\left\{J, F_1(\theta)\right\}_{+} + F_0(\theta)
\end{equation}
of the Hamiltonian of a generalized pendulum. The periodic
functions $F_i(\theta)$ depend on the angle via the kick potential
$v(\theta)$. In the lowest approximation
\begin{equation}\label{F1} F_2(\theta) = 1;\,\,\, F_1(\theta) =
-\frac{\kappa}{2} v';\,\,\, F_0(\theta) = \kappa v
+\frac{\kappa^2}{12} (v')^2\,,
\end{equation}
when the next correction
adds
\begin{equation}\label{F1^2}
\delta F_2(\theta) = -\frac{\kappa}{6} v;\,\,\, \delta
F_1(\theta) =  \frac{\kappa^2}{12} v'v;\,\,\, \delta F_0(\theta) =
-\frac{\kappa^3}{60} (v')^2 v-\frac{\kappa^3}{48} v.
\end{equation}
In further corrections higher powers of the operator $J$ also
arise.

Uprising of the angular momentum operator $J$ in (\ref{apH1})
drastically changes the eigenvalue problem. The angle $\theta$ is a
quantum-mechanical coordinate operator in this problem, and the
spectrum of the quasi-Hamiltonian ${\cal H}_1$ as well as of the
Floquet operator becomes discrete because of the periodic boundary
condition. The main effect of the term quadratic in $J$ consists in
cutting off the unrestricted kinetic energy growth. This is well
seen in Fig.1 Already the lowest correction (dotted line) describes
reasonably good the turn off from the quadratic resonant growth, as
well as the mean height of the saturation plateau. The next one
(solid line) substantially improves the description and reproduces
well also the details of quantum fluctuations in the plateau
region. Influence of the further corrections, which contain, in
particular, higher powers of the angular momentum $J$, remains weak
in the finite domain $\Delta\kappa$, the width of the resonance,
where the angular momentum $J$ in the plateau region is still
sufficiently small.

Before analyzing the conditions of validity of eq. (\ref{apH1}) in
more detail, we consider the resonance $q=2$ because of a tangible
similarity of the two cases. For $q=2$, the rotation operator
$U_r=e^{-i\pi p {\hat m}^2}$, where $p$ is an odd number, has only
two eigenvalues 1 and -1. Obviously, any periodic function
$\psi_{+}(\theta)\,\left(\psi_{-}(\theta)\right)$ which contains
only even (odd) harmonics is an eigenfunction belonging to the
eigenvalue 1 (-1). An arbitrary state $\psi(\theta)$ can be written
down as the linear superposition $\psi=\psi_{+}+\psi_{-}$ of these
eigenfunctions. Therefore, in the Hilbert space of the periodic
functions the operator $U_r$ is isomorphic to the $2\times 2$ Pauli
matrix $\sigma_3$. One can actualize this isomorphism by
representing the state $\psi(\theta)$ in the form of a
two-component spinor
\begin{equation}\label{Psi2}
\Psi(\theta) =
\begin{pmatrix}
  \psi_{+}(\theta) \\
  \psi_{-}(\theta)
\end{pmatrix}\,.
\end{equation}
In this representation the differentiation operator ${\hat m}$
which does not change harmonics' numbers grows into the diagonal
matrix
\begin{equation}\label{M}
\mathbb{M} =
\begin{pmatrix}
  {\hat m} & 0 \\
  0 & {\hat m}
\end{pmatrix}\,,
\end{equation}
while any periodic coordinate operator $F(\theta)$ looks as
\begin{equation}\label{F}
\mathbb{F}(\theta)= f_{+}(\theta)\,I +
f_{-}(\theta)\,\sigma_1
\end{equation}
with $I$ being the $2\times 2$ unity matrix.  At last, matrix
elements of dynamical operators take the form
\begin{equation}\label{me}
\mathbb{O}_{2,1}=\int_{-\pi}^{+\pi} d\theta
\Psi_2^{\dag}(\theta)\,\mathbb{O}(\theta,\mathbb{M})\,\Psi_1(\theta)\,;
\qquad \int_{-\pi}^{+\pi} d\theta \Psi^{\dag}(\theta)\,\Psi(\theta) = 1.
\end{equation}

The free rotation is described now by the matrix operator
$\mathbb{U}_r=\sigma_3 = e^{-i\frac{\pi}{2}(\sigma_3-1)}$ and the
kick operator reads as $\mathbb{U}_k=e^{-iv\sigma_1}$. Simple
manipulations with Pauli matrices lead to the following expression
for the resonant Floquet operator
\begin{equation}\label{U2}
\mathbb{U}_{p,2}^{(res)} = e^{i\frac{\pi}{2}}\,
{\textstyle
\exp\left(-i\frac{\pi}{2}\, {\bf n}\cdot\mbox{\boldmath $\sigma$}\right)}
\end{equation}
where the unit vector ${\bf n}=(0, \sin v, \cos v)$. Up to the trivial phase factor,
this is a spin-flip operator which belongs to the unitary unimodular group
$SU(2)$. Therefore, corresponding evolution in the continuous time $t$ fully reduces
to the spin rotation.

In particular, evolution of the angular momentum is given by the equation
\begin{equation}\label{M2(t)}
\Delta \mathbb{M}(t) = {\textstyle \exp\left(i\frac{\pi}{2}\,
{\bf n}\cdot\mbox{\boldmath $\sigma$}\,t\right)\, \mathbb{M}
\exp\left(-i\frac{\pi}{2}\, {\bf n}\cdot\mbox{\boldmath $\sigma$}\,t\right)\,
-\mathbb{M}}  = \Delta {\bf M}(t)\cdot\mbox{\boldmath $\sigma$}.
\end{equation}
The vector $\Delta {\bf M}(t)$
is easily calculated by making use of the formula
$$\exp\left(-i\frac{\pi}{2}\, {\bf n}\cdot\mbox{\boldmath
$\sigma$}\,t\right)= \cos(\pi t/2)-i {\bf n}\cdot\mbox{\boldmath
$\sigma$}\sin(\pi t/2).$$
Simple transformations lead to the result
\begin{equation}\label{DM2}
\Delta {\bf M}(t) = -v'\left[{\bf s}\,\sin(\pi
t/2) \cos(\pi t/2) + {\bf l}\,\sin^2(\pi t/2)\right]
\end{equation}
where the unit vectors ${\bf s}$ and ${\bf l}$ are defined by
\begin{equation}\label{uvec}
{\bf n'}=v'(0,\cos v, -\sin v)\equiv v'{\bf s};\quad {\bf l}=\left[{\bf s}\times
{\bf n}\right]=(1, 0, 0).
\end{equation}
The three unit  vectors ${\bf n}, {\bf l}, {\bf s}$ form an
orthogonal basis in the 3-dimensional adjoint space of the $SU(2)$
group. The evolution (\ref{DM2}) is purely periodic in time, so
that the kinetic energy
\begin{equation}\label{varE2}
E(t) = \frac{1}{2}\langle\left[\Delta {\bf
M}(t)\right]^2\rangle = \frac{1}{2}\langle (v')^2 \rangle\sin^2(\pi t/2) =
\frac{k^2}{4}\sin^2(\pi t/2)
\end{equation}
does not grow but rather jumps between two values $0$ and $k^2/4$
when the time $t$ runs over integer values \cite{IzSh79}.

The motion in a neighborhood of the considered resonance is
described by the quasi-Hamiltonian matrix
\begin{equation}\label{cal H2}
\mathcal{H}_2 =
\kappa\frac{\pi}{2}{\bf n}\cdot\mbox{\boldmath $\sigma$} + \kappa^2 Q(\kappa)
\end{equation}
(compare with eq. (\ref{cal H1})) where $Q$ is a $2\times 2$-matrix
operator in the spinor space. This operator satisfies the condition
(\ref{Q1}) with ${\hat m}$ substituted by the matrix $\mathbb{M}$
from eq.~(\ref{M}). With the same accuracy as above, the
quasi-Hamiltonian reads as in eq. (\ref{apH1}) again where the
angular momentum $J$ and the functions (\ref{F1}) are replaced by
$2\times2$ matrices. In the first approximation they are equal to
\begin{equation}\label{F2}
\mathbb{F}_2(\theta)=I;\,\,\,\mathbb{F}_1(\theta)=-\frac{\kappa}{2} v'\left({\bf l}+
\frac{\pi}{2}{\bf s}\right)\cdot\mbox{\boldmath $\sigma$};\,\,\,
\mathbb{F}_0(\theta)=\kappa\frac{\pi}{2}{\bf n}\cdot\mbox{\boldmath $\sigma$}+
\frac{\kappa^2}{4} ( v')^2\,I.
\end{equation}
The Hamiltonian (\ref{apH1}), (\ref{F2}) describes the motion of a
quantum particle with the spin $1/2$ along a circle in an
inhomogeneous magnetic field. The terms linear in the angular
momentum $J$ mimic a sort of the ``spin-orbital" interaction. Fig.2
(points) shows that quantitative agreement of this approximation
with exact numerical simulations (circles) worsens rather fast.
However, the second correction which can be represented in the
compact form
\begin{equation}\label{Q^1_2} \kappa^3
Q^{(1)}=\frac{\kappa}{16}\left\{J, \left\{J,
\left(\frac{\pi}{4}(v')^2{\bf n}-v{\bf l}\right)
\cdot\mbox{\boldmath $\sigma$}\right\}_{+}\right\}_{+}+
\frac{5\kappa^2}{32}\left\{J,
v'v\right\}_{+}+\frac{\kappa^3}{32}(v')^2v \left(\frac{\pi}{2}{\bf
s}-{\bf l}\right)\cdot\mbox{\boldmath $\sigma$}
\end{equation}
noticeably improves the correspondence (crosses). The two branches
correspond to even (starts from zero) and odd (starts from $k^2/4$)
kicks.

At the first glance, an important difference is seen in the $v$-dependence
of the functions $F_i$ in eq.(\ref{F1}) on the one hand side and in eq.(\ref{F2}) on
the other. In the former case, the number of harmonics do not exceeds the power of
the detuning $\kappa$. This property holds also in the higher approximations.
Likewise, the factors $k$ and $\kappa$ are balanced in the similar way so that $k$
always combines with $\kappa$ into the {\it effective} Chirikov's parameter $K_e=\kappa
k$. Afterwards only positive extra powers of $\kappa$ may remain. Therefore, the
influence of the higher terms of the expansion can be expected to be weak when
$K_e<1$.

In contrast, the unit vectors ${\bf n}$ and ${\bf s}$ in
eq.(\ref{F2}) in the most interesting case $k\gg 1$ contain $\sim
k$ harmonics and their derivatives with respect to $\theta$ are
large. This leads to terms with extra powers of $k$ in the higher
corrections, which are not compensated by the small detuning
$\kappa$ and enhance higher corrections. For example the term in
(\ref{Q^1_2}), which is quadratic with respect to the operator $J$,
contains uncompensated factor $k$. However, such terms turn out to
be inefficient  and cancel finally out due to the identity
\begin{equation}\label{U^2=1}
\left[\mathbb{U}_{p,2}^{(res)}\right]^2 =
-\exp\left(-i\pi\, {\bf n}\cdot\mbox{\boldmath
$\sigma$}\right) = I,
\end{equation}
so that the case $q=2$ does not essentially differ from the main
resonance as it concerns the role of higher corrections. This fact
can be proven by disentangling the resonant part from the squared
Floquet operator $\mathbb{U}^2_{p,2}(\kappa)$ after which the
non-diagonal $\sigma$-matrices disappear from the
quasi-Hamiltonian.

However, it is appreciably simpler to get rid of the non-trivial
$SU(2)$ algebra by separating evolution at only even and only odd
kicks. It is then enough to smooth directly the squared Floquet
transformation. Taking into account that
in the Hilbert space of periodic functions $e^{-i\pi {\hat
m}^2}\cos\theta\, e^{-i\pi {\hat m}^2}=-\cos\theta$, one comes to
the condition
\begin{equation}\label{db H2}
\mathbb{U}_{p,2}^2(\kappa) =
\exp\left(-\frac{i}{2}\kappa {\hat m}^2\right)\,
\exp\left[-\frac{i}{2}\kappa ({\hat m}-v')^2\right] =
\exp\left[-2\frac{i}{\kappa}\overline{\cal H}_2(\kappa)\right].
\end{equation}
The quasi-Hamiltonian is now found with the help of the Baker-Hausdorf
expansion,
\begin{equation}\label{db apH2}
\overline{\cal H}_2 = \frac{1}{2}({\cal K}+\bar{{\cal K}})-
\frac{i}{4\kappa}\left[{\cal K}, \bar{{\cal K}}\right]_{-}-
\frac{1}{24\kappa^2}\left[{\cal K}-\bar{{\cal K}},
\left[{\cal K}, \bar{{\cal K}}\right]_{-}\right]_{-}-
\frac{i}{48\kappa^3}
\left[{\cal K}, \left[\bar{{\cal K}},
\left[{\cal K}, \bar{{\cal K}}\right]_{-}\right]_{-}\right]_{-}+...\,.
\end{equation}
Here ${\cal K}=\frac{1}{2}J^2$ and $\bar{{\cal
K}}=\frac{1}{2}(J-\kappa v')^2= \frac{1}{2}(J+\kappa
k\sin\theta)^2$ are the kinetic energy operators at the moments
$t=0$ and $t=1$ respectively. Since each commutator gives at least
one power of the small detuning $\kappa$, uncompensated factors $k$
do not appear in the series. All four terms displayed in eq.
(\ref{db apH2}) are easily calculated explicitly though
corresponding expressions are too lengthy for presenting them here.
Fig. 3 demonstrates very good agreement of the evolution described
by this quasi-Hamiltonian with exact numerical simulations. Only
the even branch is shown.

The $q=2$ resonance of QKR is an example of the specific regimes of
motion of periodically driven systems which are known as {\it
anti-resonances}. The main feature of them is periodic exact
recurrence (see eq. (\ref{U^2=1})) after a certain number of kicks.
General consideration of the motion near anti-resonances is
presented in \cite{DES95}. The authors showed, in particular, that
such a motion has regular classical limit. In fact, this is valid
not only for $q=2$ anti-resonance but also for the actual resonance
$q=1$ (see next subsection). More than that, we will demonstrate in
secs. 4,~5  that  domains of a regular quantum motion exist near
all resonances with $q\ll l$. However, contrary to the two boundary
resonances, the widths of other resonances diminish in the
classical limit so that this motion has no direct counterpart in
the classical standard map.

\subsection{The Classical Limit}

In the both cases already considered the quantum fluctuations fade
away when the parameter $k$ increases. The quasi-Hamiltonians
${\cal H}_{1}$ and $\overline{\cal H}_{2}$ appreciably simplify in
the limit $k\gg 1, \kappa\ll 1; \,K_e~=\kappa k=const$ and the
functions $F_i(\theta)$ reduce to
\[F_2(\theta) = 1-\frac{K_e}{6}\cos\theta;\,\,\,
F_1(\theta)=\frac{K_e}{2}\sin\theta -
\frac{K_e^2}{24}\sin2\theta\,,\]
\begin{equation}\label{clF1}
F_0(\theta)=K_e\left(1-\frac{K_e^2}{240}\right)\cos\theta-
\frac{K_e^2}{24}\cos2\theta+\frac{K_e^3}{240}\cos3\theta
\end{equation}
in the case $q=1$ and to
\begin{eqnarray}
\nonumber
F_2(\theta) &=& 1+\frac{K_e}{8}-\frac{K_e}{2}\left(1+
\frac{5K_e^2}{48}\right)\cos\theta+
\frac{K_e^2}{24}\cos2\theta
-\frac{K_e^3}{32}\cos3\theta\,,\\
\label{clF2}
F_1(\theta) &=& \frac{K_e}{2}\left(1+\frac{5K_e^2}{48}\right)\sin\theta-
\frac{K_e^2}{8}\sin2\theta+
\frac{K_e^3}{96}\sin3\theta-
\frac{K_e^4}{96}\sin4\theta\,, \\
F_0(\theta) &=& -\frac{K_e^2}{8}\cos2\theta-
\frac{K_e^4}{192}\cos4\theta\,, \nonumber
\end{eqnarray}
when $q=2$. The Planck's constant disappear and corresponding
quasi-Hamiltonians pass into classical Hamilton functions which
depend on the only {\it effective} parameter $K_e$. Treating $K_e$
as the classical Chirikov's parameter, these functions coincide
with those obtained in ref. \cite{Sok67} and describe the phase
oscillations near the non-linear resonances respectively of the
first and second harmonics in the classical standard map. Fig.4
illustrates on the example of the resonance q=2 the quantum -
classics correspondence. The solid line is obtained by averaging
1000 classical trajectories with $J(t=0)=0$ over isotropic initial
angular distribution. Crosses and circles show the results of
numerical simulations of the exact quantum QKR map for two
different values of the kick parameter $k$. The effective classical
parameter is kept fixed, $K_e=0.1$. Being notably substantial for
$k=10$, the deviations due to quantum effects become inessential
when $k=100$. Agreement at large times are improved by taking into
account terms of higher powers in $J$ in the expansion of the
frequency of the non-linear phase oscillations.

The effective parameter $K_e$ differs from the product $kT$ by the
resonant part $(4\pi p/q)k$. In this connection, the suggestion
made in ref. \cite{Shp87} is worthy mentioning that in the regime
of quantum diffusion the classical Chirikov's parameter is
re-scaled as ${\tilde K}_e=2k\sin(T/2)$. This suggestion proved to
be in good correspondence with numerical simulations. In the main
order with respect to the small detuning $\kappa$ the re-scaled
value ${\tilde K}_e$ is equivalent to our $K_e$.  However, it is
not quite clear whether the whole sine make sense in the domains of
the regular motion. The quantum corrections to different structures
in the quasi-Hamiltonians have different forms neither of which can
be identified with the terms of expansion of ${\tilde K}_e$ over
the detuning $\kappa$.

We see that for both resonances considered the domains of regular
motion is estimated by the same inequality $K_e < 1$, which insures
the regular phase oscillations near corresponding non-linear
classical resonances.

Note in conclusion of this section that near the resonances $q=1,
2$ the motion does not depend on the integer number $p$. This is a
special manifestation of the following general properties of the
standard quantum map. First of all, in virtue of the identity
$e^{\pm 2\pi i {\hat m}^2}\Rightarrow 1$ the rotation operator
$U_r$ and, consequently, the Floquet transformation (\ref{U}) are
periodic in the parameter $\varsigma$ with period 1. Therefore it
is enough to restrict oneself to consideration of the interval
$0<\varsigma<1$. In reality, only half of this interval exhausts
all independent possibilities whilst the mean kinetic energy
$E(t)$ is calculated \cite{Shp87}. Indeed, presuming that
$\varsigma=1/2+\delta\varsigma;\,\,|\delta\varsigma|\leq 1/2$, one
can easily see that the transformation $\delta\varsigma\rightarrow
-\delta\varsigma;\,\, \theta\rightarrow \pi+\theta$ is equivalent
to the complex conjugation of the operator $U$ and therefore does
not change $E(t)$. Consequently, the problem investigated is
symmetric in $\varsigma$ with respect to the point $\varsigma=1/2$
and the lowest resonances $q=1, 2$ correspond to the ends of the
principal interval $0\leq\varsigma\leq 1/2$. That is why we refer
to these resonances as the boundary ones.

\section{Lowest Resonances with no Classical Limit}

Contrary to the two boundary resonances of the previous section,
those of them which lie inside the principal domain
$0<\varsigma<1/2$ have no well defined limits when
$k\rightarrow\infty,\,K_e=const$. We consider in this section the
next two ones, $q=3, 4;\,(p=1)$. They provide typical though still
exactly solvable examples of the motion near a quantum resonance.
The resonance $q=3$ is linked to the group $SU(3)$ whereas the
other one is still belongs to the same simplest group $SU(2)$ as in
the case $q=2$. Indeed, it is easy to see that the rotation
operator $U_r=e^{-i\frac{\pi}{2}{\hat m}^2}$ is equivalent to the
transformation $e^{i\frac{\pi}{4}(\sigma_3-1)}$ of the spinor
(\ref{Psi2}) \cite{ftnt}. Because of especially simple structure of
this group we start with consideration of the resonance $q=4$.

Using well known properties of the Pauli matrices we easily find for
the resonant Floquet transformation
\begin{equation}\label{U4}
\mathbb{U}_{1,4}^{(res)}=\exp\left(-iw\,{\bf n}\cdot\mbox{\boldmath $\sigma$}\right).
\end{equation}
Again, we omit a trivial phase factor of no importance. The
periodic functions in the exponent are now defined as
\begin{equation}\label{w4n}
w(v)=\arccos\left(\frac{\cos v}{\sqrt{2}}\right),\,\,\, {\bf n}(v)=\frac{1}{\sqrt{1+
\sin^2v}}(\sin v,-\sin v,-\cos v).
\end{equation}
The function $w(v)$ is the most important new element in comparison
to the $q=2$ resonance, which yields a linearly increasing term in
the angular momentum evolution
\begin{equation}\label{DM4}
\Delta {\bf M}(t) = -w'{\bf n}\,\,t -\sqrt{2}
\frac{v'}{1+\sin^2v}\left[{\bf s}\,\sin(wt) \cos(wt) +{\bf
l}\,\sin^2(wt)\right]\,.
\end{equation}
Here
\begin{equation}\label{s,l} {\bf s}=
\frac{1}{\sqrt{1+\sin^2v}}\,\frac{1}{\sqrt{2}}(\cos v,-\cos v,2\sin v), \,\,\,
{\bf l}=\left[{\bf s}\times{\bf n}\right]=\frac{1}{\sqrt{2}}(1, 1, 0).
\end{equation}
As before, the prime denotes differentiating with respect to the
angle $\theta$; the three unit vectors ${\bf n}, {\bf s}$ and ${\bf
l}$ are pairwise orthogonal. As long as the angle $\theta$ remains
fixed, the contribution of the spin rotation is periodic. However,
on the last step averaging over the angle has to be done which
leads to
\begin{equation}\label{varE4}
E(t) = r\,t^2 + \chi(t)
\end{equation}
with
\begin{equation}\label{r4}
r(k) = \frac{1}{2} \langle(w')^2\rangle = \frac{k^2}{4\pi}\int_{-\pi}^{\pi}
d\theta\sin^2\theta\frac{\sin^2v}{1+\sin^2v}= \frac{k^2}{4}\left(1-\frac{4}{\pi}
\int_0^1 dz\frac{\sqrt{1-z^2}}{1+\sin^2(k z)}\right)\,,
\end{equation}
and
\begin{equation}\label{chi4}
\chi(t) = \frac{k^2}{2\pi}\int_{-\pi}^{\pi} d\theta \sin^2\theta
\frac{\sin^2wt}{(1+\sin^2v)^2}= \frac{2k^2}{\pi}\int_0^1 dz\sqrt{1-z^2}\frac{\sin^2
\left[w(k z)\,t\right]}{\left[1+\sin^2(k z)\right]^2}\,.
\end{equation}
The function $\chi(t)$ fluctuates with time slowly approaching the
value
\begin{equation}\label{chinf}
\chi_{\infty}(k)=\frac{k^2}{\pi}\int_0^1 dz \frac{\sqrt{1-z^2}}
{\left[1+\sin^2(kz)\right]^2}\,.
\end{equation}
After extracting the constant part, the function $\chi(t)$
naturally stratifies into four smooth branches ($s=0, 1, 2,....$)
\begin{equation}\label{brchi}
\chi(t)=\chi_{\infty}(k)+\left\{
\begin{array}{rl} -\chi^{(+)}(t) & \quad t=4s, \\
\chi^{(-)}(t) & \quad t=4s+1, \\
\chi^{(+)}(t) & \quad t=4s+2,
\\ -\chi^{(-)}(t) & \quad t=4s+3.
\end{array} \right.
\end{equation}
Here
\begin{equation}\label{pmlgt}
\chi^{(+)}(t)=\frac{k^2}{4\pi}\int_0^{\pi} dv\frac{\cos 2{\tilde w}\,t}
{(1+\sin^2v)^2};\,\,\, \chi^{(-)}(t)=\frac{k^2}{4\pi}\int_0^{\pi} dv
\frac{\sin 2{\tilde w}\,t} {(1+\sin^2v)^2}
\end{equation}
where the function ${\tilde w}(v)=\pi/4-\arcsin(\cos v/\sqrt{2})$
is $2\pi$-periodic with respect to $v$ and changes from zero at
$v=0$ to the maximal value $\pi/2$ when $v=\pi$. Stationary phase
calculation gives for $\chi^{(\pm)}(t)$ the asymtotics
$const/\sqrt{t}$. Fig.5 presents an example of the function
$\chi(t)$ taken at integer values of $t$.

Simplification is possible in some limiting cases. It is easy to
see that $r(k)\approx k^4/16$ \cite{IzSh79} and
$\chi_{\infty}(k)\approx k^2/4$ when the parameter $k\ll 1$. In
this limit $\tilde{w}(kz)\approx \pi/4+k^2z^2/2$ so that
\begin{eqnarray}
\nonumber
\chi^{(+)}(t) &=& \frac{k^2}{4}\left[\cos(k^2t/2)\,J_0(k^2t/2)+
\sin(k^2t/2)\,J_1(k^2t/2)\right]\\
\label{pmchi}
\chi^{(-)}(t) &=& \frac{k^2}{4}\left[\sin(k^2t/2)\,J_0(k^2t/2)-
\cos(k^2t/2)\,J_1(k^2t/2)\right]\,.
\end{eqnarray}
The symbol $J_{\nu}(x)$ stands for the Bessel function. For small
times $k^2t\ll 1$ we find
\begin{equation}\label{pmsmt}
\chi^{(+)}(t)\approx \frac{k^2}{4}\left(1-\frac{k^4}{16}\,t^2\right),\,\,\,
\chi^{(-)}(t)\approx \frac{k^4}{16}\,t\,,
\end{equation}
while
\begin{equation}\label{lgt}
\chi^{(+)}(t)=\chi^{(-)}(t)\approx \frac{k}{4}
\sqrt{\frac{2}{\pi t}}\rightarrow 0
\end{equation}
if the time is large. In the most interesting case $k\gg 1$
simplification is achieved by averaging the fast varying factors in
the integrals over $z$ in eqs. (\ref{r4}), (\ref{chi4}). This gives
$r(k)\approx (\sqrt{2}-1)k^2/4\sqrt{2}$ which is in good agreement
with numerical data though somewhat differs from the value $k^2/12$
given in \cite{IzSh79}. At last, $\chi_{\infty}(k)\approx
3k^2/16\sqrt{2}$ in this limit. One sees that the $q=4$ resonance
admits of very detailed analytical description.

Near the resonance calculation of the lowest correction $Q^{(0)}$
gives
\[\mathbb{F}_2=I;\,\,\,
\mathbb{F}_1=-\frac{\kappa}{2}v'\left\{n_1{\bf
n}+\sqrt{2}(n_1^2+n_3^2) \left[w{\bf s}+(1-n_3w){\bf
l}\right]\right\}\cdot\mbox{\boldmath $\sigma$};\]
\begin{equation}\label{F4}
\mathbb{F}_0=\kappa w\, {\bf n}\cdot\mbox{\boldmath $\sigma$}+
\frac{\kappa^2}{2}(v')^2\left[\frac{1}{6}n_1^2+(n_1^2+n_3^2)^2(1-n_3w)\right]
\end{equation}
(compare with eq. (\ref{F2})). Computation of the next correction
$Q^{(1)}$, though making no principle problems, turns out to be rather
tedious and leads to quite cumbersome expressions. The most
important contribution comes from the correction to the matrix
$\mathbb{F}_2$
\begin{equation}
\delta {\mathbb{F}}_2(\theta)=-\frac{1}{4}(1-n_1^2)(v')^4 (w \mathbf{f}_1
+\frac{1}{6}\mathbf{f}_2)\cdot\mbox{\boldmath $\sigma$}
\end{equation}
with the vectors $\mathbf{f}_{1,2}$ given by
\[\mathbf{f}_1 = \left (n_1(3-5n_1^2)(1-n_1^2),1+2n_1^2-5n_1^4,
1+2n_1^2-5n_1^4 \right )\,,\]
\[\mathbf{f}_2 = \left (n_1n_3 (14-15n_1^2),-n_1n_3(4+15n_1^2),
-(4+7n_1^2-30n_1^4) \right)\,.\] We drop here corrections to other
matrices $\mathbb{F}$ whose influence is negligibly weak. In Fig.6
the evolution generated by the quasi-Hamiltonian (solid line) is
compared to the simulation of the exact QKR quantum map which is
shown by points. The detuning is chosen to be about $1/2$ of its
critical value after which the regime of regular motion breaks into
diffusion. The energy $E(t)$ scales with the first power of the
detuning $\kappa$ in this case. It is due to the form of the
zero-order resonant interaction which contains spin and has no
classical limit. The theory nicely reproduces all details of the
evolution up to very large times.

The width of the resonance is now much narrower than in the case of
the boundary resonances. Indeed, $\theta$-derivatives of the
functions $w(v)$ and ${\bf n}(v)$ appear in higher corrections,
which are large if $k\gg 1$. Validity of the expansion deteriorates
because of such contributions. One can roughly estimate the width
suggesting that the influence of the second order correction should
be relatively weak inside the resonant domain. This gives
$\Delta\kappa\propto k^{-2}$ which agrees with our numerical data.
Outside this interval the expansion transparently diverges.
Contrary to the boundary resonances, the region of regular motion
vanishes in the classical limit $k\rightarrow\infty$ even if the
condition $K_e=\kappa k=const$ holds.

To explore the regularity domain and adjacent area in more detail,
we have fitted in Fig.7a exact numerical data for the mean height
$E_{pl}$ of the plateau as a function of the detuning $\kappa$. Two
qualitatively different regions are clearly seen: the regularity
domain $(\kappa\leq 10^{-4})$, where the plateau is inversely
proportional to the detuning, and the quantum chaos area
$(\kappa\geq 10^{-3})$ where the height is scattered around the
generic value $l^2\sim k^4$. In the intermediate domain the higher
corrections become increasingly important and the perturbation
expansion fails. In Fig.7b similar numerical results are displayed
for the case $q=3$.

At the resonance point $q=3$ the wave function is naturally split
into three independent parts,
$\psi(\theta)=\psi_1(\theta)+\psi_2(\theta)+ \psi_3(\theta)$, each
item $\psi_\mu(\theta)= \sum_{s=-\infty}^{\infty}
C_{3s+\mu}\exp[i(3s+\mu)\theta]$ being an eigenfunction of the
rotation operator $U_r=\exp[-(2\pi i/3){\hat m}^2]$. Arranging the
items in the form of a 3-component spinor (compare with eq.
(\ref{Psi2}))
\begin{equation}\label{Psi3}
\Psi(\theta) =
\begin{pmatrix}
  \psi_1(\theta) \\
  \psi_2(\theta) \\
  \psi_3(\theta)\,,
\end{pmatrix}
\end{equation}
the rotation matrix acquires the form (we set again $p=1$)
\begin{equation}\label{Ur3}
\mathbb{U}_r=
\begin{pmatrix}
  \beta^* & 0 & 0 \\
  0 & \beta^* & 0 \\
  0 & 0 & 1
\end{pmatrix}
=\exp\left\{-\frac{2\pi i}{3}
\begin{pmatrix}
  1 & 0 & 0 \\
  0 & 1 & 0 \\
  0 & 0 & 0
\end{pmatrix}
\right\}\Rightarrow \exp\left(-\frac{2\pi i}{3}\,
\frac{1}{\sqrt{3}}\lambda_8\right).
\end{equation}
On the last step we dropped a trivial phase factor. The diagonal matrix
$\lambda_8$ is one of the standard generators of the group $SU(3)$ and
$\beta=\exp(2\pi i/3)=-1/2+i\sqrt{3}/2$. Now, since the factor $e^{\pm
i\theta}$ changes the index $\mu$ by $\pm 1$, we have in such a representation
\begin{equation}\label{v3}
e^{\pm i\theta}\Rightarrow e^{\pm i\theta}\lambda_{\pm}; \quad
\lambda_{+}=\lambda_{-}^{\dag}=
\begin{pmatrix}
  0 & 0 & 1 \\
  1 & 0 & 0 \\
  0 & 1 & 0
\end{pmatrix};
\quad \lambda_{+}\lambda_{-}=I.
\end{equation}
The matrix $\lambda_{+}$ shifts each element in the column
(\ref{Psi3}) by one position down and puts the lowest component on
the very top when the matrix $\lambda_{-}$ is the reciprocal
transformation. Both the matrices are traceless. Therefore, the
kick operator gets the form
\begin{equation}\label{kick3}
\mathbb{U}_k = \exp\left[-i\frac{k}{2}(e^{i\theta}\lambda_{+} +
e^{-i\theta}\lambda_{-})\right] =
\exp\left\{-\frac{i}{2}\left[v\,(\lambda_{+}+\lambda_{-}) -
iv'\,(\lambda_{+}-\lambda_{-})\right]\right\}.
\end{equation}
In terms of the standard $SU(3)$ generators the matrix
$\lambda_{+}$ reads
\begin{equation}\label{lmbda}
\lambda_{+}=\frac{1}{2}(\lambda_1+\lambda_4+\lambda_6)-
\frac{i}{2}(\lambda_2-\lambda_5+\lambda_7)\,.
\end{equation}
The commuting matrices $\lambda_{\pm}$ are simultaneously diagonalized,
\begin{equation}\label{diag3}
\lambda_{+}^{(diag)}=
\begin{pmatrix}
  \beta & 0 & 0 \\
  0 & \beta^* & 0 \\
  0 & 0 & 1
\end{pmatrix},
\quad
\lambda_{-}^{(diag)}=
\begin{pmatrix}
  \beta^* & 0 & 0 \\
  0 & \beta & 0 \\
  0 & 0 & 1
\end{pmatrix},
\end{equation}
with the unitary transformation
\begin{equation}\label{Phik3}
\Phi_k = \frac{1}{\sqrt{3}}
\begin{pmatrix}
  \beta^* & \beta & 1 \\
  \beta & \beta^* & 1 \\
  1 & 1 & 1
\end{pmatrix}.
\end{equation}
Correspondingly, the diagonal form of the kick is
\begin{equation}\label{dgkick3}
\mathbb{U}_k^{(diag)}=diag\, (e^{-i v_{+}}, e^{-i v_{-}}, e^{-i v_0}); \quad
v_{\pm}=k\cos(\theta\pm 2\pi/3);\,\,\, v_0=v=k\cos\theta.
\end{equation}
Obviously, $v_{+}+v_{-}+v_{0}=0$.

The resonant Floquet operator is now represented as
\begin{equation}\label{tildU3}
\mathbb{U}_{1,3}^{(res)}=
\Phi_k {\tilde \mathbb{U}}_{1,3}^{(res)} \Phi_k^{\dag}
\end{equation}
where
\begin{equation}\label{intU3}
{\tilde \mathbb{U}}_{1,3}^{(res)}= \Phi_k^{\dag}\mathbb{U}_r
\Phi_k \mathbb{U}_k^{(diag)}=
\beta^*\left[I-(1-\beta){\bf I}\cdot {\bf I}^T\right]\,U_k^{(diag)}
\end{equation}
with the one-column matrix ${\bf I}$ being equal to
\begin{equation}\label{bfI}
{\bf I} = \frac{1}{\sqrt{3}}
\begin{pmatrix}
  1 \\
  1 \\
  1
\end{pmatrix}.
\end{equation}
Due to the factorized form (\ref{intU3}) of the matrix
${\tilde\mathbb{U}}_{1,3}^{(res)}$, it can be easily diagonalized.
The eigenvectors prove to be equal to
\begin{equation}\label{tldphi3}
\tilde{\mbox{\boldmath $\phi$}}^{(\mu)}=(1-\beta)
\left[\sum_{\nu}\frac{1}{|e^{-i v_{\nu}}-u_{\mu}|^2}\right]^{-1/2}
\frac{\mathbb{U}_k^{(diag)}}{\mathbb{U}_k^{(diag)}-u_{\mu}}\,{\bf I};\quad
\mu, \nu=+,-,0
\end{equation}
while the eigenvalues $u_{\mu}=e^{-i\tilde{\epsilon}_{\mu}(\theta)}$ satisfy the
cubic equation
\begin{equation}\label{eps3}
1-(1-\beta)\,{\bf I}^T\frac{\mathbb{U}_k^{(diag)}}{\mathbb{U}_k^{(diag)}-u}{\bf I}=
1-\frac{1-\beta}{3}\sum_{\mu}\frac{e^{-i v_{\mu}}}{e^{-i v_{\mu}}-u}=0.
\end{equation}
After this equation has been solved, the resonant Floquet
transformation is represented as
\begin{equation}\label{U3}
\mathbb{U}_{1,3}^{(res)}=\exp\left(-i\Phi\tilde{\epsilon}\Phi^{\dag}\right)=
\exp\left[-iw(\theta)\,{\bf n}(\theta)\cdot\mbox{\boldmath $\lambda$}\right]; \quad
\tilde{\epsilon}= diag (\tilde{\epsilon}_{+},
\tilde{\epsilon}_{-},\tilde{\epsilon}_0)
\end{equation}
in terms of the unitary matrix $\Phi=\Phi_k\,\tilde{\Phi}$ where the tilded
matrix
\begin{equation}\label{Phi3}
\tilde{\Phi}=\left( \tilde{\mbox{\boldmath $\phi$}}^{(+)}, \tilde{\mbox{\boldmath
$\phi$}}^{(-)}, \tilde{\mbox{\boldmath $\phi$}}^{(0)}\right)
\end{equation}
consists of the eigenvectors (\ref{tldphi3}). On the last stage we
expressed the $3\times 3$ hermitian matrix in the exponent in
eq.(\ref{U3}) in terms of the $SU(3)$ generators $\lambda_a$. The
coefficients are given by
\begin{equation}\label{w;n}
w\,n_a=\frac{1}{2}tr\left[\lambda_a\,\Phi\tilde{\epsilon}\Phi^{\dag}\right];\,\,\,
a=1, 2, ..., 8.
\end{equation}
The vector ${\bf n}$ is a unit vector in the 8-dimensional adjoint
space. The transformations described give complete solution of the
problem for $q=3$ resonance. However, the final expressions in
terms of the roots of the cubic equation (\ref{eps3}) turn out to
be too cumbersome and we do not cite them here. In the similar way,
analytical solution can be found also for the resonance $q=8$ which
is linked to the $SU(4)$ group. However, algebraic problems
increase very rapidly with $q$.

Qualitatively, the situation near the lowest resonance does not
essentially differ from that near the resonance $q=4$. Figs. 8a,b
demonstrate the transition from the regular motion to diffusion
near the resonances $q=3$ and $q=4$. In both cases the resonant
growth is seen at the first stage, which then changes by a sharp
turnover. Further evolution crucially depends on the detuning. If
it is below some critical value, the asymptotic plateau begins
immediately after the turnover. However, above this value the
plateau is preceded by the stage of diffusion. It is clearly seen
that the slope at the latter stage ($\kappa \cdot E\propto t$) is
twice as smaller as that at the resonant stage ($\kappa\cdot
E\propto t^2$).

\section{General Case}
\subsection{Floquet Operator for a Resonance of Arbitrary Order}

The consideration presented above can be easily extended to a resonance of an
arbitrary order $q$. The wave function is expressed in the form of a $q$-component
complex vector
\begin{equation}\label{Psiq}
\Psi^T(\theta)=\left(\psi_1(\theta)\,\, \psi_2(\theta)\,...\,\psi_q(\theta)\right).
\end{equation}
The component
\begin{equation}\label{item}
\psi_\mu=\sum_{s=-\infty}^{\infty} C_{sq+\mu}\,\exp\left[i(sq+\mu)\theta\right];
\qquad \mu=1, 2, 3, ..., q
\end{equation}
is an eigenfunction of the operator $U_r^{(p=1)}\equiv\exp(-2\pi
i{\hat m}^2/q )$, which belongs to the eigenvalue $\exp\left[(-2\pi
i/q)\, mod(\mu^2, q)\right]$. Therefore, the rotation for the
resonance $(p,q)$ is implemented by the diagonal matrix
\begin{equation}\label{Ur,pq}
\mathbb{U}_r(p,q) = \exp\left(-\frac{2\pi i p}{q}\lambda_0\right)
\end{equation}
where all matrix elements of the diagonal matrix $\lambda_0$ are
integers from the interval $[1, q]$. The kick operator looks as in
eq. (\ref{kick3}) where now
\begin{equation}\label{Kq}
\lambda_{+}=\lambda^{\dag}_{-}=
\begin{pmatrix}
  {\bf 0}^T & 1 \\
  I & {\bf 0}
\end{pmatrix}
\end{equation}
and ${\bf 0}$ is the $(q-1)$-dimensional zero column vector when
$I$ is the unit matrix of the same dimension. Both these matrices
are traceless and their properties are similar to those in the case
$q=3$ described above. In particular, they are simultaneously
diagonalized with the matrix
\begin{equation}\label{Phikq}
\left(\Phi_k^{(\mu)}\right)_{\nu}=\frac{1}{\sqrt{q}}
\left(\beta^*\right)^{\mu\nu}=\frac{1}{\sqrt{q}}
\exp\left(-\frac{2\pi i}{q}\mu\nu\right).
\end{equation}
of eigenvectors which are the discrete plane waves inside a sample
of the length $q$ in the angular momentum space. These waves
satisfy the periodic boundary conditions at the ends of the sample.
The diagonal representation of the kick operator is a natural
extension of eq. (\ref{dgkick3})
\begin{equation}\label{dgkickq}
\mathbb{U}_k^{(diag)}=diag\, \left(e^{-iv_1}, e^{-iv_2}, ..., e^{-iv_q}\right);
\quad v_{\mu}=k\cos\left(\theta+\frac{2\pi}{q}\mu\right); \quad \sum_{\mu} v_{\mu}=0.
\end{equation}
Thus, matrix elements of the kick operator are given by the finite sums
\begin{equation}\label{mkick}
\left(\mathbb{U}_k\right)_{\mu\nu}=
\frac{1}{q}\sum_{\varrho=1}^{q} e^{-\frac{2\pi i}{q}(\mu-\nu)\varrho}\,e^{-iv_{\varrho}}
\end{equation}
while  the resonant Floquet matrix reads
\begin{equation}\label{mFlq}
\left(\mathbb{U}_{p,q}^{(res)}\right)_{\mu\nu}=
e^{-\frac{2\pi ip}{q} mod(\mu^2,q)}\left(\mathbb{U}_k\right)_{\mu\nu}.
\end{equation}
Additional phases in eq. (\ref{mFlq}), which become quasi-random
when the integers $p$ and $q$ are large enough, spoil the
uniformity along the sample and disrupt the plane waves
(\ref{Phikq}). This results in localization of eigenvectors of the
resonant Floquet matrix \cite{FGP82,FFGP85,Izr90}. On the other
hand, the rotation operator is converted with the reciprocal
transformation into the matrix
\begin{equation}\label{gamma} \gamma_{\mu-\mu'}=
\sum_{\nu=1}^{q}\left(\Phi_k^{(\nu)}\right)_{\mu'}^*\,
e^{-\frac{2\pi ip}{q}\,\nu^2}\left(\Phi_k^{(\nu)}\right)_\mu=
\frac{1}{q}\sum_{\nu=1}^{q}e^{-\frac{2\pi ip}{q}\,\nu^2}\,
e^{-\frac{2\pi i}{q}\nu (\mu-\mu')}.
\end{equation}
Eqs. (\ref{dgkickq}, \ref{gamma}) bridges our representation to
that used in the ref. \cite{IzSh79}.

However, the both representations yet considered do not permit
extending to the case of a finite detuning from the resonance. To
make this possible, we have at first to represent the resonant
Floquet operator in the from of an united $SU(q)$ transformation
(compare with (\ref{U3})),
\[\mathbb{U}^{(res)}_{p,q} = e^{-\frac{2\pi ip}{q}\lambda_0}\,
\exp\left[-i\frac{k}{2}(e^{i\theta}\lambda_{+}+e^{-i\theta}
\lambda_{-})\right]=e^{-\frac{2\pi ip}{q}\lambda_0}\,
\exp\left[-i(v\lambda_{R}+v'\lambda_{I})\right]=\]
\vspace{-18pt}
\begin{equation}\label{Uq}
\Rightarrow \exp(-i\Phi\tilde{\epsilon}\Phi^{\dag})=
\exp\left(-i w\,{\bf n}\cdot\mbox{\boldmath $\lambda$}\right)
\equiv \exp\left(-i\tilde{\mathcal{H}}^{(res)}\right)
\end{equation}
where
\begin{equation}\label{R,I}
\lambda_R=\frac{1}{2}(\lambda_{+}+\lambda_{-})\,\,\,,
\lambda_I=\frac{1}{2i}(\lambda_{+}-\lambda_{-})
\end{equation}
when the matrices $\lambda_a;\,a=1, 2, ..., q^2-1$ are the standard
generators of the fundamental q-dimensional representation of the
group $SU(q)$. As before, the irrelevant phase factor generated by
the trace $tr\lambda_0$ is omitted in final expression. This
implies the condition
tr$\,\tilde{\epsilon}=\sum_\mu\tilde{\epsilon}_\mu=0$.

Diagonalization of the matrix (\ref{mFlq}) cannot be carried out
analytically if the dimension $q$ of the fundamental representation
exceeds 4. Nevertheless, some generic conclusions can be drawn from
eq. (\ref{Uq}) even without explicit knowing the functions
$w(\theta)$ and $\mathbf{n}(\theta)$. The diagonal eigenvalue
(quasienergy) matrix $\tilde{\epsilon}$ is connected to the them as
\begin{equation}\label{epsUp}
\tilde{\epsilon}=w\left(\mathbf{n}\cdot\Phi^{\dag}\mbox{\boldmath
$\lambda$}\Phi\right)
\end{equation}
where $\Phi$ is the unitary matrix
\begin{equation}\label{Phiq}
\Phi=\left(\mbox{\boldmath $\phi$}^{(1)},\,\mbox{\boldmath $\phi$}^{(2)},\,...,\,
\mbox{\boldmath $\phi$}^{(q)}\right)
\end{equation}
of the normalized eigenvectors of the resonant Floquet operator.
Matrix elements of the $q^2-1$ matrices $\Phi^{\dag}\lambda_a\Phi$
form in $(q^2-1)$-dimensional adjoint space $q^2$ vectors
$\mbox{\boldmath $\Upsilon$}^{(\mu,\nu)}$ with the components
\begin{equation}\label{Ups}
{\mbox{\boldmath $\phi$}^{(\mu)}}^{\dag}\lambda_a
\mbox{\boldmath $\phi$}^{(\nu)}\equiv \sqrt{2}\Upsilon_a^{(\mu,\nu)}.
\end{equation}
The diagonal part of (\ref{epsUp}) reads in these terms
\begin{equation}\label{tileps}
\tilde{\epsilon}_\mu=\sqrt{2}w\left(\mathbf{n}\cdot\mbox{\boldmath
$\Upsilon$}^{(\mu)}\right)\,,\quad
\mbox{\boldmath $\Upsilon$}^{(\mu)}\equiv \mbox{\boldmath
$\Upsilon$}^{(\mu,\mu)}\,,
\end{equation}
whereas the off-diagonal part yields the orthogonality condition
\begin{equation}\label{nortUp}
\mathbf{n}\cdot\mbox{\boldmath $\Upsilon$}^{(\mu,\nu)}=0;
\,\,\,\mu\neq\nu\,.
\end{equation}
Because of hermiticity of the generators $\lambda_a$, all $q$
vectors $\mbox{\boldmath $\Upsilon$}^{(\mu)}$ are real. For the
same reason the remaining vectors with $\mu\neq\nu$ make up
$q(q-1)/2$ mutually complex conjugate pairs $\mbox{\boldmath
$\Upsilon$}^{(\mu,\nu)}$, $\mbox{\boldmath
$\Upsilon$}^{(\nu,\mu)}={\mbox{\boldmath $\Upsilon$}^{(\mu,\nu)}}^*$.
Using the known property
\begin{equation}\label{lamlam}
\sum_a\left(\lambda_a\right)_{\rho\sigma}\left(\lambda_a\right)_{\tau\upsilon}
=2\left(\delta_{\rho\upsilon}\delta_{\sigma\tau}-
\frac{1}{q}\delta_{\rho\upsilon}\delta_{\sigma\tau}\right)
\end{equation}
of generators of the $SU(q)$ group, we easily find for scalar
products of the vectors $\mbox{\boldmath $\Upsilon$}$
\begin{equation}\label{sprod}
{\mbox{\boldmath $\Upsilon$}^{(\mu,\nu)}}^*\cdot\mbox{\boldmath $\Upsilon$}^{(\mu',\nu')}
=\left(\delta_{\mu\mu'}\delta_{\nu\nu'}-
\frac{1}{q}\delta_{\mu\nu}\delta_{\mu'\nu'}\right).
\end{equation}
In particular, any given vector $\mbox{\boldmath
$\Upsilon$}^{(\mu,\nu)}$ with $\mu\neq\nu$ is orthogonal to all
others including the vectors $\mbox{\boldmath $\Upsilon$}^{(\mu)}$.
Therefore, the set $\{\mbox{\boldmath
$\Upsilon$}^{(\mu,\nu)};\,\,\mu\neq\nu\}$ constitutes an
orthonormalized basis in the $q(q-1)$-dimensional subspace of the
adjoint space. As to the vectors $\mbox{\boldmath
$\Upsilon$}^{(\mu)}$ which lie in the complementary orthogonal
$(q-1)$-~dimensional subspace, they are neither orthogonal,
\begin{equation}\label{nort}
\mbox{\boldmath $\Upsilon$}^{(\mu)}\cdot\mbox{\boldmath $\Upsilon$}^{(\mu')}
=\left(\delta_{\mu\mu'}-\frac{1}{q}\right)
\end{equation}
(see (\ref{sprod})), nor linearly independent because of relation
\begin{equation}\label{lrel}
\sum_\mu \mbox{\boldmath $\Upsilon$}^{(\mu)}=
{\rm tr}\mbox{\boldmath $\lambda$}=0\,.
\end{equation}
Excluding with the help of this relation one of the vectors, we can
use the rest of them to construct an orthonormalized basis in the
complementary subspace as well. In fact, the overfullness mentioned
becomes insignificant in the limit of large $q$.

On the other hand, it follows from (\ref{epsUp}) that
\begin{equation}\label{Upeps}
w\mathbf{n}=\frac{1}{2}{\rm tr}\left(\tilde{\epsilon}\,
\Phi^{\dag}\mbox{\boldmath $\lambda$}\Phi\right)=
\frac{1}{\sqrt{2}}\sum_{\mu}\tilde{\epsilon}_\mu
\mbox{\boldmath $\Upsilon$}^{(\mu)}.
\end{equation}
Using also eq. (\ref{nort}) together with the fact that
$\sum_\mu\tilde{\epsilon}_\mu=0$ we obtain finally
\begin{equation}\label{w,n}
w=\frac{1}{\sqrt{2}}\left(\sum_{\mu}
\tilde{\epsilon}_\mu^2\right)^{\frac{1}{2}};
\qquad
\mathbf{n}=\frac{1}{\sqrt{2}\,w}\sum_{\mu}\tilde{\epsilon}_\mu
\mbox{\boldmath $\Upsilon$}^{(\mu)}.
\end{equation}
Note that the vector $\mathbf{n}$ entirely belongs to the
$(q-1)$-dimensional subspace spanned by the vectors $\mbox{\boldmath
$\Upsilon$}^{(\mu)}$.

\subsection{Resonant Evolution}

In general case we have similar to eq. (\ref{M2(t)})
\[\Delta\mathbb{M}(t)=e^{i\tilde{\mathcal{H}}^{(res)}\,t}
\left[\mathbb{M}\,,\,e^{-i\tilde{\mathcal{H}}^{(res)}\,
t}\right]_{-}=
-\Phi\left(\frac{d\tilde{\epsilon}}{d\theta}\,t+ie^{i\tilde{\epsilon} t}
\left[\Phi^{\dag}\frac{d\Phi}{d\theta}\,,\,e^{-i\tilde{\epsilon} t}
\right]_{-}\right)\Phi^{\dag}\]
\vspace{-18pt}
\begin{equation}\label{Mq(t)}
=\Delta\mathbf{M}(t)\cdot\mbox{\boldmath $\lambda$}
\equiv \left[\mathbf{M}^{(0)}\,t+\mathbf{M}^{(1)}(t)\right]
\cdot\mbox{\boldmath $\lambda$}\,.
\end{equation}
The equation obtained is nothing but the matrix form of the result
of the paper \cite{IzSh79}.

One can advance as follows. First of all we find from this equation
\begin{equation}\label{Mq0}
\mathbf{M}^{(0)}=-\frac{1}{2}{\rm tr}\left(\frac{d\tilde{\epsilon}}
{d\theta}\,\Phi^{\dag}\mbox{\boldmath $\lambda$}\Phi\right)=
-\frac{1}{\sqrt{2}}\sum_\mu\tilde{\epsilon}'_\mu\mbox{\boldmath
$\Upsilon$}^{(\mu)}\,,
\end{equation}
and
\[\mathbf{M}^{(1)}(t)=-\frac{i}{2}{\rm tr}\left(e^{i\tilde{\epsilon}
t} \left[\Phi^{\dag}\frac{d\Phi}{d\theta}\,,\,e^{-i\tilde{\epsilon}
t} \right]_{-}\Phi^{\dag}\mbox{\boldmath
$\lambda$}^{(\mu)}\Phi\right)\]
\vspace{-18pt}
\begin{equation}\label{Mq1}
=-\frac{i}{\sqrt{2}}\sum_{\mu,\nu}
\left[e^{i(\tilde{\epsilon}_\mu-\tilde{\epsilon}_\nu)t}-1\right]
\left({\mbox{\boldmath $\phi$}^{(\mu)}}^*\cdot
\frac{d\mbox{\boldmath $\phi$}^{(\nu)}}{d\theta}\right)
{\mbox{\boldmath $\Upsilon$}^{(\mu,\nu)}}^*\,.
\end{equation}
Terms with $\mu=\nu$ drop out the latter sum. Therefore, the two
vectors $\mathbf{M}^{(0)}$ and $\mathbf{M}^{(1)}$ lie in the
mutually orthogonal subspaces. The vector $\mathbf{M}^{(1)}(t)$ is
a quasi-periodic function of time for any fixed value of the angle
$\theta$. On the other hand, direct calculation gives at the moment
$t=1$
\begin{equation}\label{DMq1}
\Delta\mathbb{M}(1)=\mathbb{U}_k^{\dag}\left[\mathbb{M}\,,
\mathbb{U}_k\right]_{-}=
-v'\lambda_R+v\lambda_I\equiv
\mathbf{u}(v,v')\cdot\mbox{\boldmath $\lambda$}\,.
\end{equation}
Here we took into account that the free rotation operator
$\mathbb{U}_r^{(res)}$ commutes with the angular momentum
$\mathbb{M}$ and that the kick operator reads
$\mathbb{U}_k=\exp\left[-i(v\lambda_{R}+v'\lambda_{I})\right]$ (see
eq. (\ref{Uq})). Components of the real vector ${\bf u}(v,v')$ are
easily calculated as
\begin{equation}\label{V_a}
u_a=\frac{1}{2}v\,{\rm tr}(\lambda_a\lambda_I)-
\frac{1}{2}v'\,{\rm tr}(\lambda_a\lambda_R)\,.
\end{equation}
Comparison with eqs. (\ref{Mq(t)})-(\ref{Mq1}) taken at the same
moment $t=1$ gives
\begin{equation}\label{Mq-u}
{\tilde{\epsilon}'_\mu}=-\sqrt{2}\,\mathbf{u}\cdot\mbox{\boldmath
$\Upsilon$}^{(\mu)}\,,\quad
i\left({\mbox{\boldmath $\phi$}^{(\mu)}}^*\cdot
\frac{d\mbox{\boldmath $\phi$}^{(\nu)}}{d\theta}\right)=
-\sqrt{2}\frac{\mathbf{u}\cdot\mbox{\boldmath $\Upsilon$}^{(\mu,\nu)}}
{e^{i(\tilde{\epsilon}_\mu-\tilde{\epsilon}_\nu)}-1}\,,
\end{equation}
so we come to
\begin{equation}\label{Mq0-u}
\mathbf{M}^{(0)}=\sum_\mu \left(\mathbf{u}\cdot\mbox{\boldmath
$\Upsilon$}^{(\mu)}\right)\,\mbox{\boldmath $\Upsilon$}^{(\mu)}
\end{equation}
and
\begin{equation}\label{Mq1-u}
\mathbf{M}^{(1)}(t)=\sum_{\mu\neq\nu}\frac{1-e^{i(\tilde{\epsilon}_\mu
-\tilde{\epsilon}_\nu) t}}{1-e^{i(\tilde{\epsilon}_\mu-\tilde{\epsilon}_\nu)}}
\left(\mathbf{u}\cdot\mbox{\boldmath
$\Upsilon$}^{(\mu,\nu)}\right)\,
{\mbox{\boldmath $\Upsilon$}^{(\mu,\nu)}}^*
\end{equation}
instead of eqs. (\ref{Mq0}) and (\ref{Mq1}). All derivatives
disappear from these expressions.

There exists another representation of the evolution which is more
suitable for our further purpose. It comes from the connection
\[\Delta\mathbb{M}(t)=
-\int_0^t d\tau\, e^{i\tilde{\mathcal{H}}^{(res)}\,\tau}\,
\left(\frac{d}{d\theta}\tilde{\mathcal{H}}^{(res)}\right)
e^{-i\tilde{\mathcal{H}}^{(res)}\,\tau}\,\]
\vspace{-18pt}
\begin{equation}\label{EqM(t)}
=-w'\left({\bf n}\cdot\mbox{\boldmath $\lambda$}\right)\,t
-w\,{\mathbf{n}'}^T\int_0^t d\tau\,e^{-i \tilde{\mathcal{L}}^{(res)}\tau}\,
\mbox{\boldmath $\lambda$}\,.
\end{equation}
Here
\begin{equation}\label{Lv}
\tilde{\mathcal{L}}^{(res)}=w\,\mathbf{n}\cdot\mbox{\boldmath $\Lambda$}
\end{equation}
and the matrix relation
\begin{equation}\label{f-adj}
\exp\left(i w\,{\bf n}\cdot\mbox{\boldmath $\lambda$}\,\tau\right)\,
\left(\mathbf{a}\cdot\mbox{\boldmath $\lambda$}\right)
\exp\left(-i w\,{\bf n}\cdot\mbox{\boldmath $\lambda$}\,\tau\right)\,
=\mathbf{a}^T\exp\left(-i w\,{\bf n}\cdot\mbox{\boldmath
$\Lambda$}\,\tau\right)\mbox{\boldmath $\lambda$}\,,
\end{equation}
where $\mathbf{a}$ is a $(q^2-1)$-dimensional vector, has been used.
The matrices $\Lambda_a$ whose matrix elements look as
\begin{equation}\label{mLam}
(\Lambda_a)_{bc}=-2if_{abc},
\end{equation}
with the quantities $f_{abc}$ being the structure constants of
$SU(q)$ group, are generators of the adjoint representation of the
group. From dynamical point of view, the interrelation of the
operators $\tilde{\mathcal{H}}^{(res)}=w({\bf n}\cdot
\mbox{\boldmath $\lambda$})$ and
$\tilde{\mathcal{L}}^{(res)}=w({\bf n}\cdot\mbox{\boldmath
$\Lambda$})$ is that between Hamilton and Liouville operators.

Now, let the vector $\mbox{\boldmath $\chi$}^{(a)}$ be an
eigenvector which belongs to the eigenvalue $\tilde{l}_a$ of the
Liouville matrix $\tilde{\mathcal{L}}^{(res)}$. In accordance with
the meaning of the adjoint representation, this vector obeys the
condition
\begin{equation}\label{eichi}
\bigl[{\bf n}\cdot\mbox{\boldmath $\lambda$}\,,
\mbox{\boldmath $\chi$}^{(a)}\cdot\mbox{\boldmath
$\lambda$}\bigr]_{-}=\tilde{l}_a\mbox{\boldmath $\chi$}^{(a)}
\cdot\mbox{\boldmath $\lambda$}\,.
\end{equation}
As one can easily convince oneself by direct substitution, the
vectors $\mbox{\boldmath $\Upsilon$}^{(\mu,\nu)}$ satisfy the
equation of the form (\ref{eichi}) with the eigenvalues
$\tilde{\epsilon}_\nu-\tilde{\epsilon}_\mu$. This elucidates the
meaning of the vectors $\mbox{\boldmath $\Upsilon$}$ as the
eigenmodes of the Liouville operator. It is convenient to
re-number $q(q-1)/2$ solutions with $\mu>\nu$ with the help of
the superscript $a=\beta=1,2,...,q(q-1)/2$ so that
\begin{equation}\label{Up-chi}
\mbox{\boldmath $\Upsilon$}^{(\mu,\nu)}\Rightarrow\mbox{\boldmath $\chi$}^{(\beta)};
\quad
\tilde{\epsilon}_\nu-\tilde{\epsilon}_\mu\Rightarrow\tilde{l}_{\beta}.
\end{equation}
Then for the solutions with $\nu>\mu$
\begin{equation}\label{*}
\mbox{\boldmath $\chi$}^{(-\beta)}={\mbox{\boldmath $\chi$}^{(\beta)}}^*;
\quad \tilde{l}_{-\beta}=-\tilde{l}_{\beta}.
\end{equation}
There also exist $q-1$ real eigenvectors
$\mbox{\boldmath$\chi$}^{(\alpha)}; \alpha=1,2,...,q-1$ with zero
eigenvalues $\tilde{l}_{\alpha}=0$. All these {\it zero modes} are
pairwize orthogonal linear superpositions of the vectors
$\mbox{\boldmath $\Upsilon$}^{(\mu)}$. Obviously, one of them is
just the vector $\mbox{\boldmath $\chi$}^{(1)}=\mathbf{n}$ defined
in eq. (\ref{w,n}).

Again, we can separate in eq. (\ref{EqM(t)}) the linearly growing
and quasi-periodic contributions and then exclude the derivatives
$w'$ and $\mathbf{n}'$ by comparing the result with
eq.(\ref{DMq1}). In such a way we arrive at
\begin{equation}\label{M0V}
\mathbf{M}^{(0)}=\sum_{\alpha}
\left(\mathbf{u}\cdot\mbox{\boldmath $\chi$}^{(\alpha)}\right)
\mbox{\boldmath $\chi$}^{(\alpha)}\,,
\end{equation}
\vspace{-20pt}
\begin{equation}\label{M1V}
\mathbf{M}^{(1)}(t)=
\sum_{\beta} \frac{1-e^{-il_{\beta}t}}{1-e^{-il_{\beta}}}\,
\left(\mathbf{u}\cdot\mbox{\boldmath$\chi$}^{(\beta)}\right)
{\mbox{\boldmath $\chi$}^{(\beta)}}^*\,.
\end{equation}
The result (\ref{M1V}) is just identical to (\ref{Mq1-u}). As
to eqs. (\ref{Mq0-u}) and (\ref{M1V}), they express the vector
$\mathbf{M}^{(0)}$ in terms of different sets of vectors which are
linear combinations of each other and belong to the same set of the
degenerate zero eigenvalues. Thus we come to the conclusion
that the resonant growth entirely originates from the zero
Liouville modes while the non-zero ones yield quasi-periodic
evolution.

Kinetic energy at the moment $t$ is equal to
\begin{equation}\label{varEq}
E(t)=
\frac{1}{2}\sum_{a,b}\langle\Delta M_a(t)\Delta M_b(t)\rangle
\Psi_0^{\dag}\lambda_a\lambda_b\Psi_0=
\frac{1}{2}\sum_a\eta_a
\langle\left[\Delta M_a(t)\right]^2\rangle.
\end{equation}
The symbol $\langle...\rangle$ stands for the $\theta$-averaging.
As before, the initial state $\Psi_0$ is chosen to be isotropic,
$\Psi^{\dag}= (0,0,...,1)$, so that
$\Psi_0^{\dag}\lambda_a\lambda_b\Psi_0= \delta_{ab}\eta_a$ with
\begin{equation}\label{eta}
\eta_a=0\,\, {\rm if}\, a\leq (q-1)^2-1; \quad
\eta_a=1\,\, {\rm if}\, (q-1)^2\leq a\leq q^2-2;\quad
\eta _{q^2-1}=2(1-1/q).
\end{equation}
The first set of indices enumerates such matrices $\lambda_a$ which
include the $SU(q-1)$ generators. Such matrices annul the initial
state $\Psi_0$. The second set marks $2q-1$ non-diagonal
$\lambda$-matrices with non-zero elements in the qth columns and
rows. The matrix $\lambda_q$ is the last diagonal generator of the
group $SU(q)$. So, only those modes $\mbox{\boldmath$\chi$}^{(a)}$
are significant which have appreciable projections onto the
$(2q-1)$-dimensional {\it active} subspace indicated above.

According to eq. (\ref{varEq}), the energy resonant growth rate is
is given by
\begin{equation}\label{rq}
r(p,q;k)=
\frac{1}{2}\sum_a \eta_a \langle\left[M_a^{(0)}\right]^2\rangle.
\end{equation}

Contribution of the non-zero modes fluctuates with time and
approaches asymptotically a finite positive value
\begin{equation}\label{E1inf}
\chi_{\infty}(p,q;k)=\lim_{t\rightarrow\infty}
\frac{1}{2}\sum_a \eta_a \langle\left[M_a^{(1)}(t)\right]^2\rangle\,.
\end{equation}
Estimation of the interference terms is less certain. If the
stationary points $dl_{\beta}/d\theta=0$ are not degenerate, terms
$\propto\sqrt{t}$ arise together with the linearly growing ones.

\subsection{Vicinity of a Resonance}

Motion near the point of a resonance of the order $q$ is described by the
quasi-Hamiltonian
\[\mathcal{H}=\kappa\,\tilde{\mathcal{H}}^{(res)}+\kappa^2\mathbb{Q}(\kappa)=
\kappa\,\tilde{\mathcal{H}}^{(res)}+\kappa^2\mathbb{Q}^{(0)}+
\kappa^3\mathbb{Q}^{(1)}+...\]
\vspace{-20pt}
\begin{equation}\label{cal Hq}
=\mathbb{F}_0(\theta)+
\frac{1}{2}\left\{\mathbb{J}, \mathbb{F}_1(\theta)\right\}_{+}+
\frac{1}{2}\mathbb{J}\mathbb{F}_2(\theta)\mathbb{J}+...
\end{equation}
where
\begin{equation}\label{J,F}
\mathbb{J}=I\otimes J,\quad \mathbb{F}_j(\theta)=
F_j(\theta)+\mathbf{F}_j(\theta)\cdot\mbox{\boldmath $\lambda$}\,.
\end{equation}
The operators $\mathbb{F}_j$ are found from the matrix analogs
of the conditions (\ref{eqQ1^0}), (\ref{eqQ1^1}). For the first
correction eq. (\ref{eqQ1^0}) gives $\mathbb{F}_2(\theta)=I$ while
\begin{equation}\label{eqF1q}
\int_0^1 d\tau \mathbb{F}_1(\theta; -\tau)=-\int_0^1 d\tau
\Delta\mathbb{J}(-\tau)\,,
\end{equation}
\begin{equation}\label{eqF0q}
\int_0^1 d\tau \mathbb{F}_0(\theta; -\tau)=
-\int_0^1 d\tau \left(\left[\Delta\mathbb{J}(-\tau)\right]^2+
\left\{\Delta\mathbb{J}(-\tau),
\mathbb{F}_1(\theta; -\tau)\right\}_{+}\right)\,.
\end{equation}
Here
\begin{equation}\label{defJF}
\mathbb{F}_j(\theta;-\tau)=e^{-i\mathcal{H}^{(res)}\tau}\,
\mathbb{F}_j(\theta)\,e^{i\mathcal{H}^{(res)}\tau},\quad
\Delta\mathbb{J}(-\tau)=\kappa\,\Delta\mathbb{M}(-\tau)\,.
\end{equation}
The l.h.s. in eqs. (\ref{eqF1q}), (\ref{eqF0q}) are calculated
with the help of the connection (\ref{f-adj}),
\begin{equation}\label{lhs}
\int_0^1 d\tau \mathbb{F}_j(\theta; -\tau)=F_j(\theta)+
\mathbf{F}_j^T(\theta)
\int_0^1 d\tau\,e^{i\tilde{\mathcal{L}}^{(res)}\tau}
\cdot\mbox{\boldmath $\lambda$}\,.
\end{equation}
Because of zero modes, a reciprocal to the operator
${\tilde{\mathcal{L}}^{(res)}}$ does not exists. This prohibits
from straightforward integration over $\tau$. To avoid the said
difficulty, we regularize operator
${\tilde{\mathcal{L}}_\delta^{(res)}}={\tilde{\mathcal{L}}^{(res)}}-
i\delta I$ where infinitesimal $\delta$ must be set to zero at the
very end of calculation. Then the operator
\begin{equation}\label{Intq}
\int_0^1 d\tau\,e^{i\tilde{\mathcal{L}}_\delta^{(res)}\tau}=
-\,\frac{1-e^{i\tilde{\mathcal{L}}_\delta^{(res)}}}
{i\tilde{\mathcal{L}}_\delta^{(res)}}
\end{equation}
becomes well defined. Now, in virtue of eqs. (\ref{EqM(t)}),
(\ref{M0V}) and (\ref{M1V})
\begin{equation}\label{DelJ}
\Delta\mathbb{J}(-\tau)=\kappa\,\mathbf{u}^T\,
\frac{1-e^{i\tilde{\mathcal{L}}_\delta^{(res)}\tau}}
{1-e^{-i\tilde{\mathcal{L}}_\delta^{(res)}}}
\cdot\mbox{\boldmath $\lambda$}\,.
\end{equation}
Substituting eqs. (\ref{lhs}), (\ref{Intq}) and (\ref{DelJ}) in the
condition (\ref{eqF1q}), we finally find
\begin{equation}\label{F1q}
\mathbb{F}_1(\theta)=\kappa\,\mathbf{u}^T\frac{1+
i\tilde{\mathcal{L}}_\delta^{(res)}-
e^{i\tilde{\mathcal{L}}_\delta^{(res)}}}
{\left(1-e^{i\tilde{\mathcal{L}}_\delta^{(res)}}\right)
\left(1-e^{-i\tilde{\mathcal{L}}_\delta^{(res)}}\right)}
\cdot\mbox{\boldmath $\lambda$}\,.
\end{equation}
Due to skew symmetry of the Liouville matrix,
$(\tilde{\mathcal{L}}^{(res)})^T=-\tilde{\mathcal{L}}^{(res)}$, the
matrix $\mathbb{F}_1(\theta)$ is hermitian as eq. (\ref{cal Hq})
implies. Other conditions are solved in the similar way. However,
corresponding expressions are very bulky in general case.When the
kick parameter $k\gg 1$ the typical number of harmonics in
operators $\mathbb{F}_j(\theta)$ is proportional to $k$, with a
rapidly increasing coefficient $\xi(q)$. This results in very fast
diminishing of the widths of resonances when $q$ grows.

In the approximation (\ref{cal Hq}) the quasi-Hamiltonian
$\mathcal{H}$ is formally equivalent to the Hamiltonian of a
quantum particle with $q$ intrinsic degrees of freedom which moves
in a $(q^2-1)$-component inhomogeneous ``magnetic" field. In a
sense, such a motion is a quantum analog of the classical phase
oscillations near a non-linear resonance.

\section{Convergency Problem}

Consideration presented above shows that in some domain of the
detuning from a quantum resonance of a finite order $q$ evolution
of QKR looks like a conservative motion described by the effective
time-independent quasi-Hamiltonian with a discrete quasi-energy
spectrum. A few lowest terms of the expansion (\ref{cal Hq})
allow to predict with great accuracy the evolution for a very long
time. More than that, in the range $\Delta\kappa$ of such a domain
(the width of the resonance), which is determined by the condition
that the influence of higher corrections is week, the accuracy is
improved with each further correction kept. Nevertheless, it is
clear that the formal expansion (\ref{cal Hq}) cannot converge.
Indeed, within the width of any strong resonance of a relatively
small order there exist an infinite number of resonances of large
orders which give rise to unrestricted resonant energy growth.
Independently of the number of corrections taken into account, the
quasi-Hamiltonian approach fails to reproduce such a growth which
implies continuous spectrum.

Actually, the resonant rate $r(p,q;k)$ decreases when the order $q$
increases and becomes exponentially small if $q$ noticeably exceeds
the typical localization length. In Fig.9 we plot the empirical
dependence of the resonant growth rate on the resonance order $q$.
At each $q$  the ratio $p/q$ is chosen to be as close to the "most
irrational" number $(\sqrt{ 5}-1)/2$ as possible. Our data are in
agreement with earlier results from \cite{IzSh79,T93,CFGV86}. The
solid line shows the fit with the help of the semi-empirical formula
\begin{equation}\label{Chf}
r(p,q;k)=\frac{2k^2}{3q}\exp\left(-q/2l\right)
\end{equation}
proposed in \cite{Chir99}.

The qualitative arguments presented in \cite{Chir89} connect the
exponential suppression of the resonant rate with the localization
and tunneling in the momentum space. The wave function
corresponding to an eigenvector $\mbox{\boldmath $\phi$}^{(\mu)}$
reads \cite{IzSh79}
\begin{equation}\label{ceif}
\psi^{(\mu)}(\theta)=
\sum_{m=-\infty}^{\infty} \psi^{(\mu)}_{m}\,e^{im\theta}=
\sum_{\nu=1}^q \phi_\nu^{(\mu)}(\theta_0)\,
e^{i\nu\,(\theta-\theta_0)}\frac{1}{2\pi}\sum_{s=-\infty}^{s=\infty}
e^{isq\,(\theta-\theta_0)}\,.
\end{equation}
One must distinguish here between the coordinate eigenvalue
$\theta_0$ and the argument $\theta$ of the coordinate
representation. In the angular momentum representation this
equation yields
$|\psi^{(\mu)}_{\nu+sq}|=|\phi_\nu^{(\mu)}(\theta_0)|$, so that the
angular momentum distribution is periodic with the period $q$.

On account of the quasi-random phases in the matrix (\ref{mFlq})
(see also discussion below this formula) the $q$-dimensional
eigenvectors $\mbox{\boldmath $\phi$}^{(\mu)}$ are, in fact,
localized so that the number $l$ of their appreciable components is
much smaller than $q$. Then overlap of neighboring bumps of an
eigenfunction $\psi^{(\mu)}_m$ is, typically, exponentially weak.
This resembles the eigenfunctions of a particle moving in a
periodic chain of potential wells. The resonant grows of the
angular momentum of QKR is similar to the transport through the
chain of the particle which initially was localized in some well.
The rate $r(p,q;k)$ is an analog of the mean value of the squared
group velocity of such a particle. The latter is proportional to
the exponentially small probability of tunneling between
neighboring wells. This interpretation is in agreement with
relation \cite{IzSh79}
\begin{equation}\label{rq1}
r(p,q;k)=\frac{1}{2}\sum_\mu \langle\left(\tilde{\epsilon}_\mu'\right)^2\,
|\phi^{(\mu)}_q|^2\rangle
\end{equation}
which directly follows from eq. (\ref{Mq(t)}) and contains the
weighted-mean value of the squared "group velocities"
$\tilde{\epsilon}_\mu'$. Returning to the expression (\ref{rq}), we
conclude that in the case $q\gg l$ the Liouville zero modes with
exponential accuracy lie in a subspace orthogonal to the active
one.

The aforecited arguments show that in the case of large $q$ the
resonant growth reveals itself only on a very remote time
asymptotics owing to the tunneling between localized parts of the
globally delocalized quasienergy eigenfunctions. In particular, if
such a resonance hits the domain of a strong one, it happens after
the exponentially large time
\begin{equation}\label{t-as}
t_r\approx \left[\frac{A}{\kappa\, r(p,q;k)}\right]^{1/2}
\propto \exp(q/4l)
\end{equation}
when quadratically growing contribution becomes comparable with the
height of the plateau $E_{pl}\approx A/\kappa$ formed due to the
influence of the strong resonance (see Figs 7a,b).  We found it
difficult to calculate the coefficient $A$ analytically but
numerical simulations show that under condition $k\gg 1$ it weakly
depends on the kick parameter $k$ as well as on the order of the
strong resonance.

Exponential effects of such a kind, which are characteristic
of the tunneling, are well known to be beyond the reach of
perturbation expansion. For this reason they cannot be described in
the framework of the quasi-Hamiltonian method. This approach
reproduces only those features of the motion which are determined
by the discrete component of the quasienergy spectrum. In
particular, contribution of the non-zero modes much faster attains
its asymptotic value (\ref{E1inf}). As a result, for all weak
resonances inside the width of a strong one the time dependence of
of the part $\mathbf{M}^{(1)}(t)$ is dictated during exponentially
long times $t\lesssim t_r$ by their strongest brother .

On the other hand, if the order of the resonance we are interested
in is very large, $q\gg l$, and this resonance lies in the region
of typical irrationals being far from all strong resonances,
already a very small detuning suffices for killing the quadratic
growth with exponentially small rate. At the same time, such a
shift does not influence the term $\mathbf{M}^{(1)}(t)$ which
reproduces on exponentially large (though finite) time scale all
characteristic features of the ``localized quantum chaos''
\cite{Izr90}.

The behaviour is most complicated and ambivalent in the transient
region $q\sim l$. In this case a number of resonances with
comparable and moderate orders are neighboring and their domains
can overlap. The expansion near one of them forms a plateau which
lasts until the quadratic growth in a next resonance of the same
strength reveals itself so that the original expansion fails.
However, the expansion near the new resonance cuts off the growth
and forms a higher plateau until a next resonance comes to the
action. Such a pattern of repeatedly reappearing regimes of
resonant growth has been discovered in \cite{CFGV86}.

\section{Summary}

In this paper we propose on the example of the QKR model the
concept of the time independent quasi-Hamiltonian of a quantum map.
The regimes of quantum resonances, which take place under
conditions $k=const$ and $\varsigma=T/4\pi=p/q$, play a crucial
role in our construction. The motion in the very point of a quantum
resonance with the order $q$ is, generally, exactly described by a
continuous transformation from the $SU(q)$ group. The two
qualitatively different contributions: growing with time and
saturating, in the evolution of the QKR come respectively from the
zero and non-zero modes of the generator of the $SU(q)$
transformation in the adjoint representation of the group. A
perturbation expansion exists near the point of a given quantum
resonance, which provides quite a good description of the motion
within some domain - the width of the resonance. Inside the width
of a strong resonance the motion is mastered by the resonance. This
motion is proved to be similar to that of a quantum particle with q
intrinsic degrees of freedom along a circle in an inhomogeneous
$(q^2-1)$-component ``magnetic" field.

The width of a quantum resonance strongly depends on its order $q$.
The resonances with smallest orders are the strongest ones and have
maximal widths. The motion within the width of such a resonance,
being dominated by it, is proved to be regular. In all cases save
the two boundary resonances $q=1, 2$ the regular quantum motion
exists in spite of the fact that the corresponding classical motion
is chaotic and exponentially unstable. The widths of these
resonances vanish in the classical limit
$k\rightarrow\infty\,,T\rightarrow 0\,, K_{c}=kT=const$. Such a
situation holds as long as the condition $q\ll l$ takes place,
where $l$ is the localization length for close typical irrational
$\varsigma$, although the widths rapidly diminishe with $q$. In the
opposite case of very large orders $q\gg l$ and $p\sim q$ the
motion weakly depends on the $q$ and as well as on the detuning
$\kappa$. The motions with rational and irrational $\varsigma$
differ only on a very remote time asymptotics. During a long though
finite time the motion reveals universal features characteristic
for the localized quantum chaos.

\section{Acknowledgments}

We are very grateful to B.V. Chirikov for making his results
available to us prior to publication, discussions and many
illuminating remarks, to F.M. Izrailev for many important remarks
and advises and D. Shepelyansky for critical reading of a
preliminary version of this paper. Two of us (V.V.S. and O.V.Zh.)
thank D.V. Savin for help and advises. V. Sokolov is indebted to
the International Center for the study of dynamical systems in
Como, where the reported investigation was started and to the Max
Planck Institute for the physics of complex systems in Dresden,
where the main part of the paper was written, for generous
hospitality extended to him during his stays. Financial
support from the Cariplo Foundation and the Russian Fond for
fundamental researches, grant No 99-02-16726 (V.V.S. and O.V.Zh.),
the EU program Training and Mobility of Researches, contract No
ERBFMRXCT960010 and Gobierno Autonomo de Canarias (D.A.) are
acknowledged with thanks.

\newpage
\twocolumn

\begin{figure}
%\epsfxsize 6cm
%\rotate[r]{\epsffile{fig1.ps}}
\psfig{file=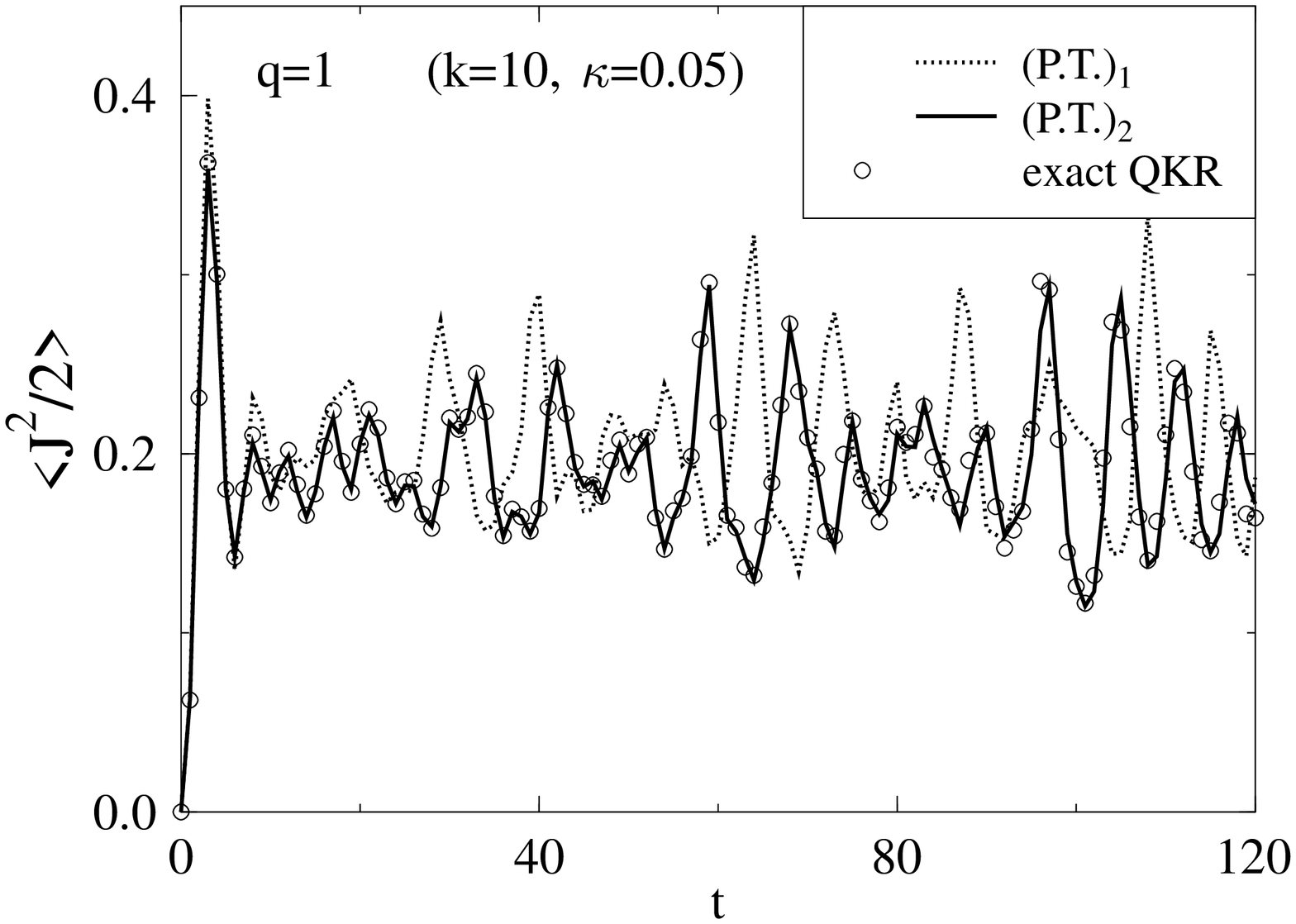,width=0.45\textwidth,rheight=0.3\textwidth,angle=-90}
\caption{ The kinetic energy $<J^2/2>$ versus the number of kicks
$t$ for the main resonance $q=1$. The dotted and solid lines show
predictions of the proposed theory in the first and second orders,
respectively. Open circles correspond to the exact QKR map.}
\label{fig1}
\end{figure}

%Fig.2
\begin{figure}
\psfig{file=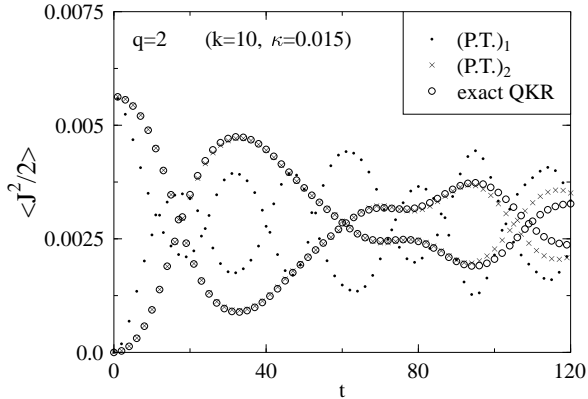,width=0.45\textwidth,rheight=0.3\textwidth,angle=-90}
\caption{The kinetic energy $<J^2/2>$ versus the number of kicks
$t$ for $q=2$. Dots and crosses show predictions of the theory
in the first and second orders. Open circles correspond to the exact
QKR map. The two (almost) symmetric branches are due to the
only {\it even} or only {\it odd} kicks respectively.}
\label{fig2}
\end{figure}

%Fig.3
\begin{figure}
\psfig{file=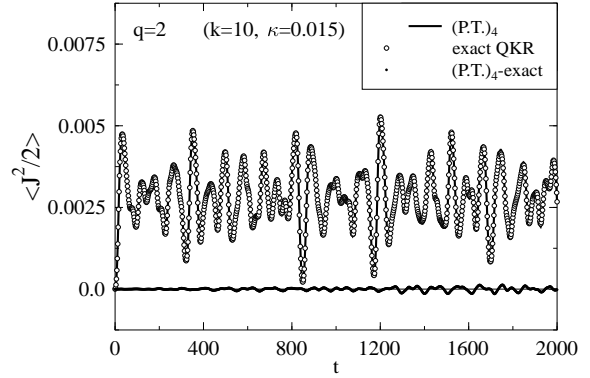,width=0.45\textwidth,rheight=0.3\textwidth,angle=-90}
\caption{The kinetic energy $<J^2/2>$ versus the number of {\it even}
kicks for $q=2$. The solid line shows evolution predicted by the
quasi-Hamiltonian (\ref{db apH2}) while the open circles correspond
to exact numerical simulations. The dotted curve at the bottom indicates
the deviation of the theory from the exact solution.}
\label{fig3}
\end{figure}

%Fig.4
\begin{figure}
\psfig{file=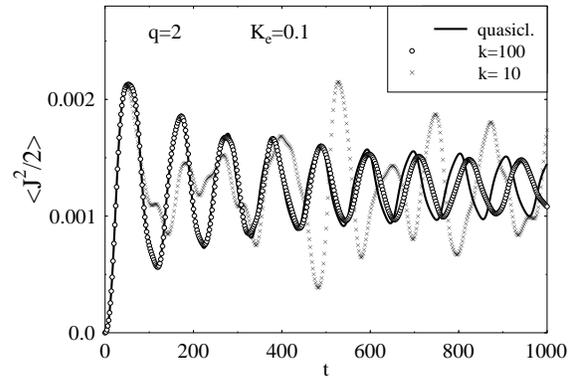,width=0.45\textwidth,rheight=0.3\textwidth,angle=-90}
\caption{The kinetic energy $<J^2/2>$ versus the number of {\it even}
kicks for $q=2$. The solid line results from the classical limit
(\ref{clF2}); crosses and open circles correspond to the exact QKR map
at the same effective classical parameter $K_e$ but different $k$.}
\label{fig4}
\end{figure}

%Fig.5
\begin{figure}
\psfig{file=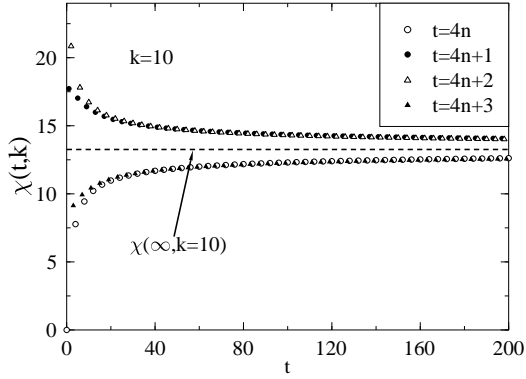,width=0.45\textwidth,rheight=0.3\textwidth,angle=-90}
\caption{The four branches of the function $\chi(t,k)$.
Asymptotically all of them converge to the same limit
$\chi(\infty,k)$ shown by dashed line.}
\label{fig5}
\end{figure}

%Fig.6
\begin{figure}
\psfig{file=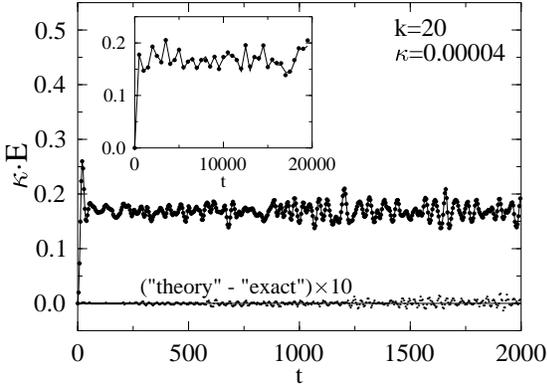,width=0.45\textwidth,rheight=0.3\textwidth,angle=-90}
\caption{The kinetic energy $E$ versus the kick number $t$ for
$q=4$. The solid line shows the second order perturbation theory
 and the points correspond to the exact QKR map (each 4th kick
 is kept in the main part and each 500th in the inset).
 Deviations of the theory from the exact map is indicated at the
 bottom. }
\label{fig6}
\end{figure}

%Fig.7a
\begin{figure}
\psfig{file=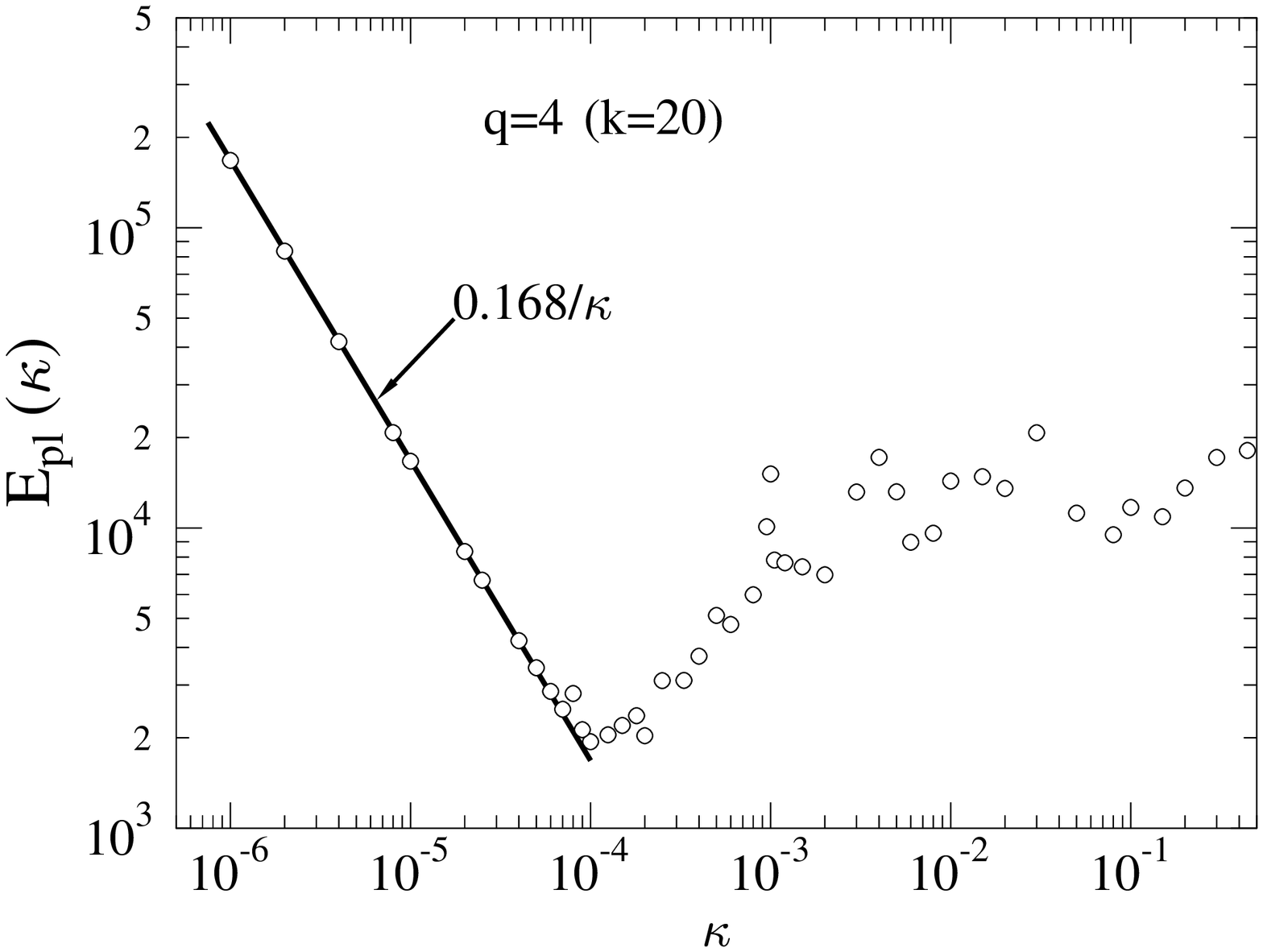,width=0.45\textwidth,rheight=0.3\textwidth,angle=-90}
\caption{The plateau height $E_pl$ versus detuning $\kappa$
 near the resonance $q=4$. The solid line corresponds to the
theoretical relation $E_pl\propto 1/\kappa$.}
\label{fig7a}
\end{figure}

%Figs.7b
\begin{figure}
\psfig{file=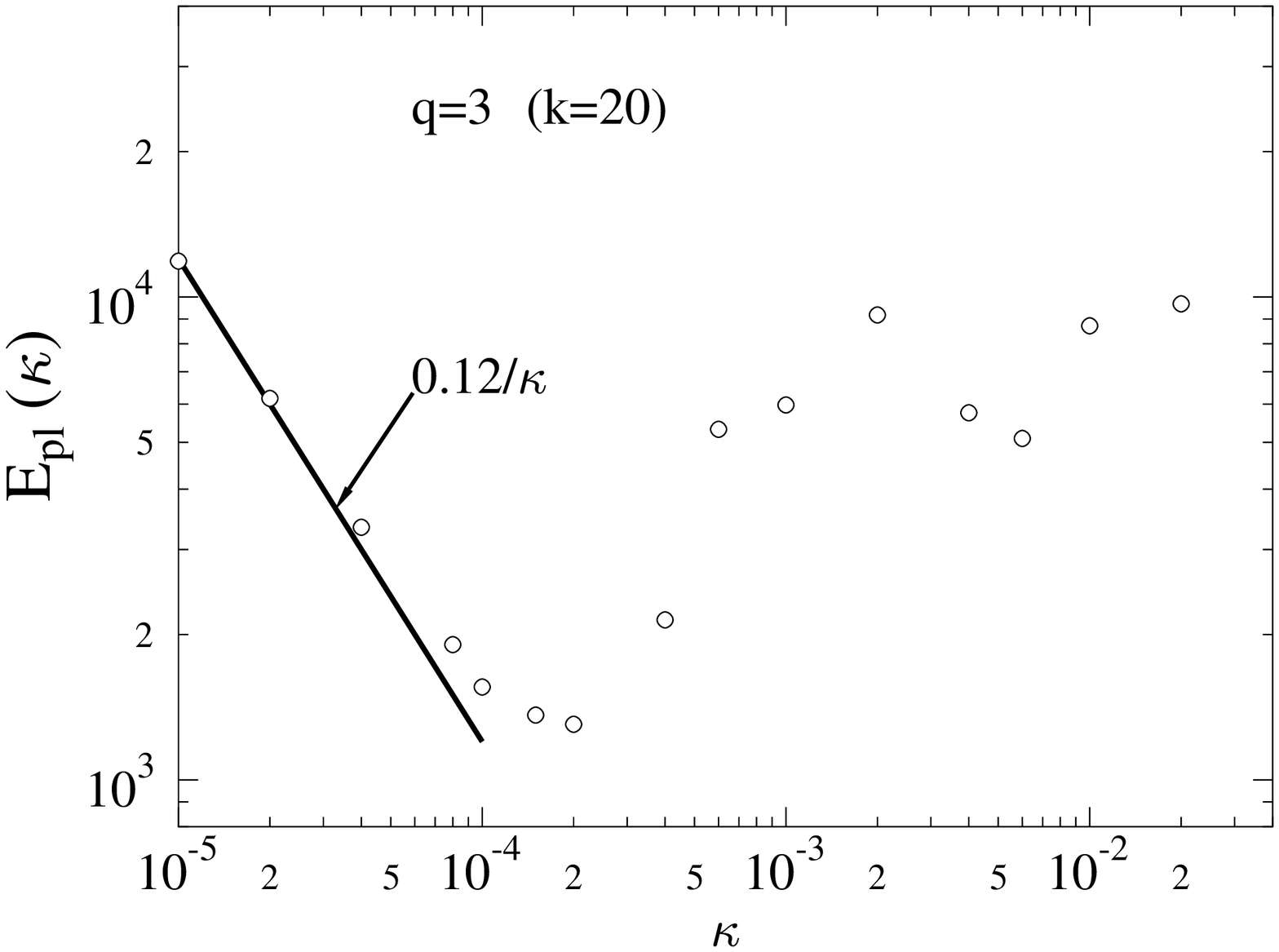,width=0.45\textwidth,rheight=0.3\textwidth,angle=-90}
\caption{The plateau height $E_pl$ versus detuning $\kappa$
 near the resonance $q=3$. The solid line corresponds to the theoretical
 relation $E_pl\propto 1/\kappa$.}
\label{fig7b}
\end{figure}

%Fig.8a,b
\begin{figure}
%\mbox{\epsfxsize 6.5cm  \rotate[r]{\epsffile{fig8a.ps}}}
\psfig{file=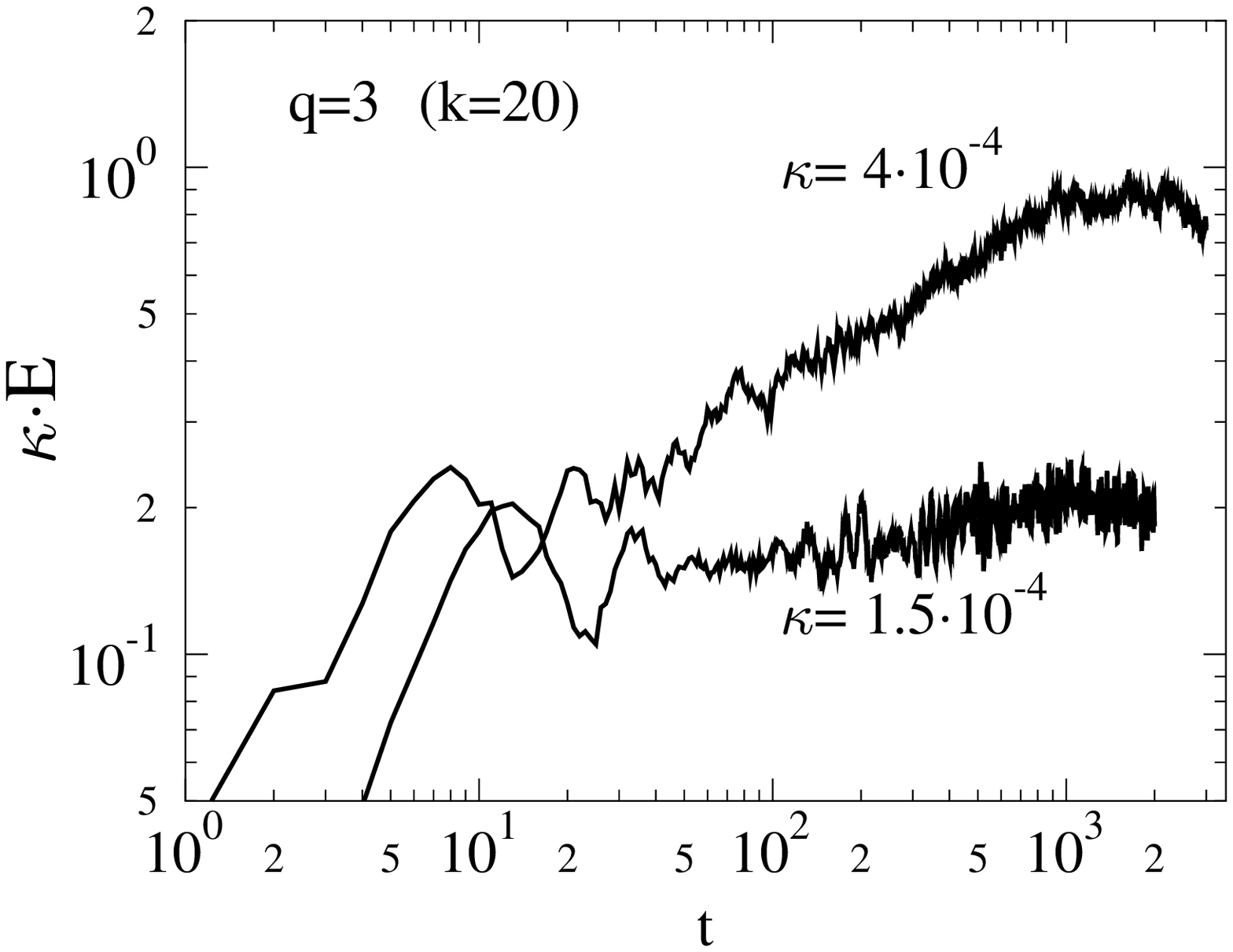,width=0.45\textwidth,rheight=0.35\textwidth,angle=-90}
%\mbox{\epsfxsize 6.5cm  \rotate[r]{\epsffile{fig8b.ps}}}
\psfig{file=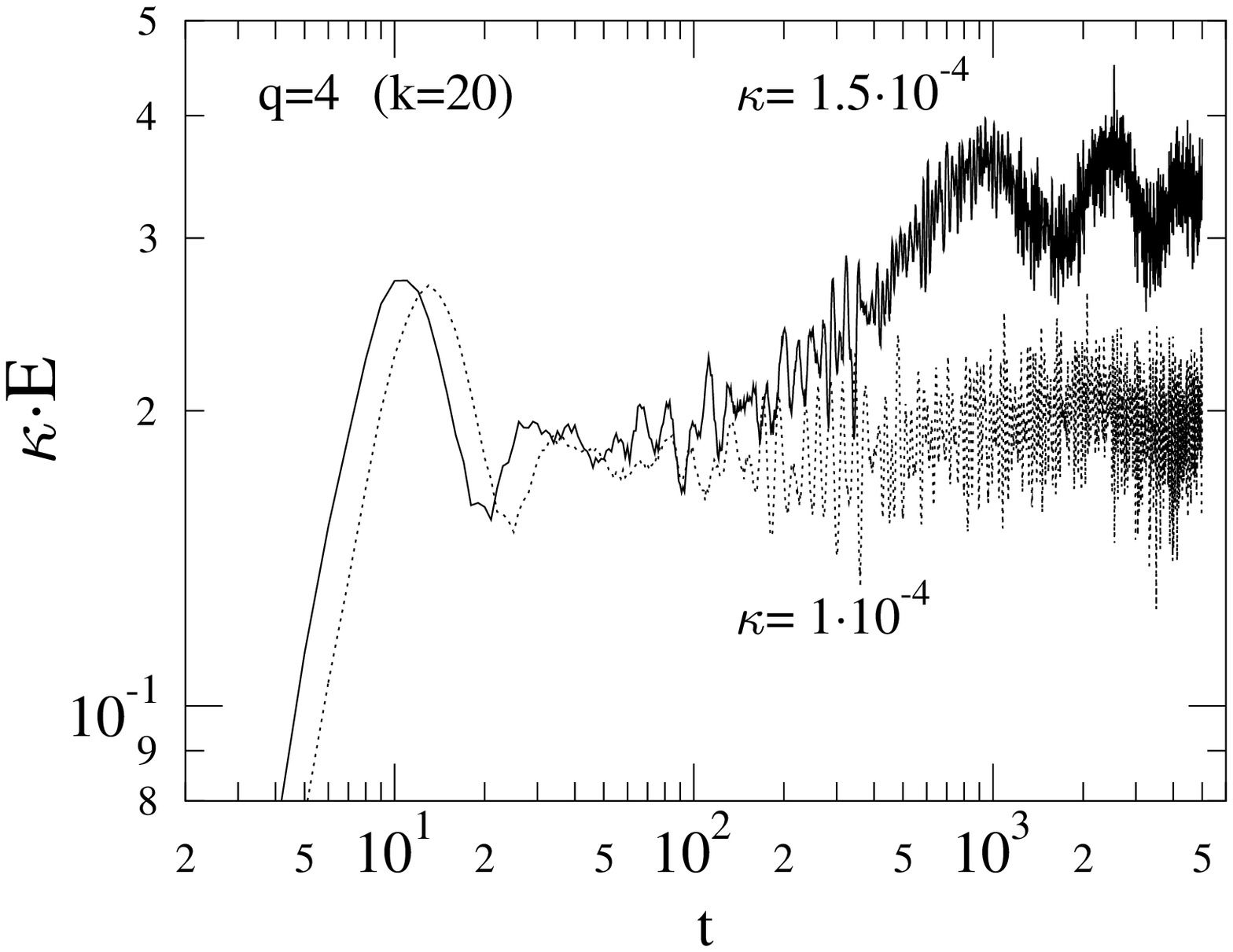,width=0.45\textwidth,rheight=0.35\textwidth,angle=-90}
\caption{The crossover region near the resonances $q=3$ and $q=4$.
The upper curves correspond to $\kappa$ outside the resonance
widths. The diffusion is clearly seen after a short initial
resonant stage while at smaller values of the detuning the
diffusion does not develop. Double log scale is chosen to show all
stages of the time evolution.}
\label{fig8}
\end{figure}

%Fig.9
\begin{figure}
\psfig{file=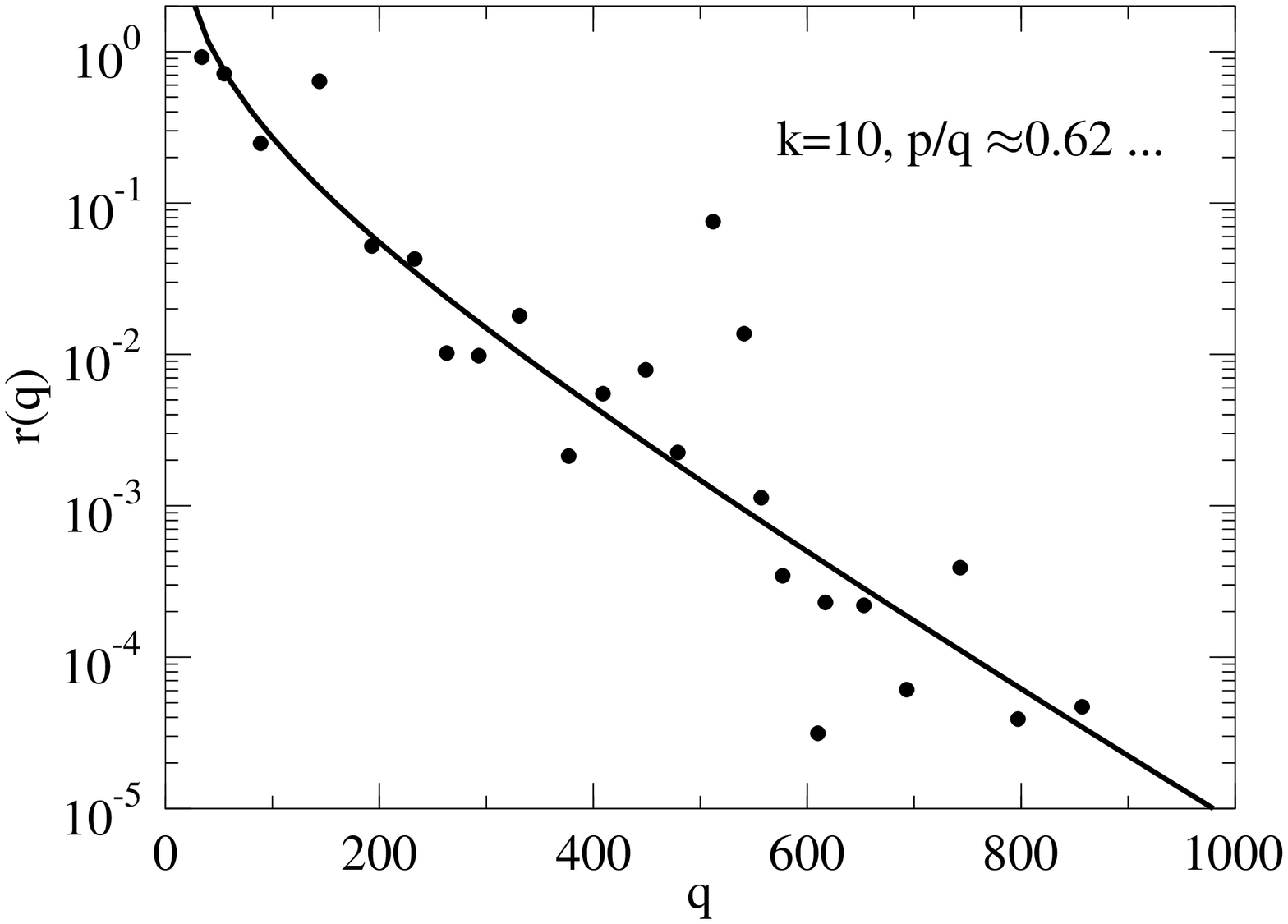,width=0.45\textwidth,rheight=0.3\textwidth,angle=-90}
\caption{The resonant growth rate $r(q)$ versus the resonance
 order $q$. At each $q$ the $p/q$ ratio is chosen to be the
 closest to the most irrational number $(\sqrt(5)-1)/2$.}
\label{fig9}
\end{figure}


\begin{thebibliography}{99}
\bibitem{ftnt}
 The matrix dimension of the resonant Floquet operator reduces to $q/2$ \\
 if $q$ is divisible by 4, see ref. \cite{IzSh79}.
\end{thebibliography}
\end{document}